\documentclass[aps,pra,notitlepage,floatfix]{revtex4-1}
\usepackage{graphics,graphicx,amssymb,subfig,array,txfonts,subfig,multirow,caption,verbatim,bm,bigstrut,color,psfrag,textcomp}
\begin{document}
\addtolength{\oddsidemargin}{-.13in}
\title{Large photon number extraction from individual atoms trapped in an optical lattice}
\author{M. D. Shotter}
\email{Email: mshotter@gmail.com}
\affiliation{National Institute of Standards and Technology,
100 Bureau Drive, Stop 8423,
Gaithersburg, MD 20899-8423, USA}
\altaffiliation{Previously: Department of Physics, University of Oxford, Oxford, UK OX1 3PU}
\date{24 November 2010}
\begin{abstract}
The atom-by-atom characterization of quantum gases requires the development of novel measurement techniques. One particularly promising new technique demonstrated in recent experiments uses strong fluorescent laser scattering from neutral atoms confined in a short-period optical lattice to measure the position of individual atoms in the sample. A crucial condition for the measurements is that atomic hopping between lattice sites must be strongly suppressed despite substantial photon recoil heating. This article models three-dimensional polarization gradient cooling of atoms trapped within a far-detuned optical lattice. The atomic dynamics are simulated using a hybrid Monte Carlo and master equation analysis in order to predict the frequency of processes which give rise to degradation or loss of the fluorescent signal during measurements. It is shown, consistent with the experimental results, that there exists a wide parameter range in which the lifetime of strongly-fluorescing isolated lattice-trapped atoms is limited by background gas collisions rather than radiative processes. In these cases the total number of scattered photons can be as large as $10^8$ per atom. The performance of the technique is related to relevant experimental parameters. 
\end{abstract}
\maketitle
\section{Introduction}
The detailed characterization of non-linear many-body quantum dynamics is a major goal of ultracold atomic and condensed matter physics. In recent years there has been significant experimental progress in creating strongly interacting quantum systems within ultracold atomic ensembles (see the review article \cite{bloch_manybody_2008} and references therein, and recent work including \cite{cheinet_counting_2008,gemelke_in_2009}). The strong interaction between neutral alkali atoms and laser radiation allows for the prospect that such strongly interacting quantum systems can be probed and investigated at the resolution of single atoms. This has been demonstrated very recently in two experiments \cite{bakr_probing_2010,sherson_single_2010}; it may be anticipated that the excellent results from these experiments will stimulate further experimentation using similar techniques. 

The motivation to measure the spatial position of each atom in a strongly interacting quantum system is the desire to probe its structure and properties at an unprecedented level of detail. Strongly interacting quantum systems tend to exhibit much more complex dynamics than weakly interacting or linear quantum systems; this complexity leads to emergent behavior such as exotic phases of the quantum matter. Such behavior is often hard to simulate and fully understand on a theoretical level. Atom-by-atom measurements on these systems would allow much more detailed cross-checking of theoretical models against real data than is currently available, facilitating a much more complete understanding of these complicated dynamics. The substantial flexibility available in choosing the system Hamiltonian has led the idea that such systems may be used as `quantum simulators' to probe various many-body Hamiltonians of interest in condensed matter physics \cite{jaksch_cold_2005,cho_condensed-matter_2008}. 

In order to take images of the atomic distribution at single-atom resolution, recent experiments \cite{nelson_imaging_2007,bakr_quantum_2009,bakr_probing_2010,sherson_single_2010} have used a deep far-detuned optical lattice to confine and mutually exclude the atoms as they are being measured. To detect the presence of an atom at a site with high confidence, the atoms must be localized to individual lattice sites throughout the duration of the measurement, despite substantial photon recoil heating. This means that the atoms need to be cooled as they are fluorescing; in current experiments polarization gradient cooling is used, with the cooling light collected to form the fluorescent signal. If atoms were to hop between lattice sites, not only would the spatial resolution of the signal be degraded, but there is a good chance that atoms would be lost from the measurement region entirely by undergoing a light-assisted collision with another atom, or hopping to the edge of the lattice. Therefore, for efficient detection, hopping between lattice sites should be minimized while fluorescent scattering should be maximized. 

The purpose of this article is to model the physical process of polarization gradient cooling of isolated atoms in a deep far-detuned optical lattice in order to understand and support future experiments performed using this technique. In particular, the dependence of the atomic hopping rate between sites of the lattice is investigated as a function of experimental parameters. 

The available literature concerned with the polarization gradient cooling of confined atoms is limited to the study of atomic dynamics in dissipative optical lattices (for a comprehensive review, see \cite{grynberg_cold_2001}), and does not directly address the situation at hand. The polarization gradient cooling of ions has also been studied in separate work \cite{cirac_laser_1992,cirac_laser_1993}. 

In common with most theoretical treatments of laser cooling, this article uses a semi-classical Monte Carlo technique in order to model polarization gradient cooling in a computationally tractable manner. The current situation differs from previous work in the context of dissipative optical lattices (e.g.\ \cite{petsas_semiclassical_1999,jonsell_nonadiabatic_2006,castin_quantization_1991}) in the use of a separate far-detuned optical lattice to confine the atoms and a higher than usual atomic saturation parameter during the cooling process; these both help to maximize the fluorescent scattering to hopping rate ratio. 

This article also differs from previous articles by the use of a novel hybrid Monte Carlo-Master Equation (HMCME) technique to extract the quantity of interest---the hopping rate between wells---from the simulation. This is necessary due to the difficulty in extracting the hopping rate from the Monte Carlo simulation alone. The difficulty is due to the great disparity in time scales; while intra-atomic dynamics take place on a timescale of $10^{-6}\,$s, it is desired that atomic hopping between lattice sites takes place on a timescale greater than $10^{2}\,$s. The direct Monte Carlo simulation of the hopping is therefore unfeasible. The approach taken uses Monte Carlo simulations of the short-time dynamics to construct a master equation rate model for the simulation of the long-time behavior. This is a novel approach for the simulation of ultracold atomic dynamics.

The present article studies the dynamics of atoms undergoing polarization gradient cooling in a deep far-detuned optical lattice. While this is the basis for optical resolved-atom measurements of the spatial distribution of an atomic sample, other considerations are also important in optical resolved-atom experiments. Light-assisted collisions quickly eject atom pairs from multiply-occupied wells during the first small fraction of the measurement period; in current experiments the parities of the site occupation numbers are measured rather than the true spatial distribution of the atoms. Another consideration is that it is desirable that the design of the optical apparatus allows the resolution of each individual lattice site within the imaging plane. Furthermore, if the atomic sample is three-dimensional, scattering from atoms in the imaging plane should be distinguishable from the scattering from out-of-focus atoms. These considerations, while important, relate to the interpretation of the measurement data and to the design of the optical apparatus; as such, they may be divorced from the basic physical process under examination in this article, and are left to be discussed elsewhere \cite{shotter_design_2010}. 

Looking beyond the two-dimensional measurements of the recent experiments \cite{bakr_probing_2010,sherson_single_2010}, three-dimensional resolved-atom tomographic measurements of atomic distributions are possible with suitable apparatus \cite{nelson_imaging_2007}. Tomographic measurements require multiple exposures to probe the complete atomic distribution; this substantially increases the time needed for the measurement. This places correspondingly stricter requirements on the atomic fluorescence and hopping rates, and it becomes much more important to optimize the scattering rate from each atom without recoil heating leading to atom loss during the measurement period.

The simulation developed in this article is used in a separate article by the author \cite{shotter_design_2010} to inform the design of an experimental method which is capable of three-dimensional tomographic measurements of the position of each atom in strongly interacting quantum systems at half-wavelength spatial resolution. The article (Ref.\ \cite{shotter_design_2010}) outlines a potential solution to the problems of resolvability and light-assisted collisions during fluorescent measurements of dense ultracold atomic systems. Furthermore, a solution to the out-of-plane scattering problem---the background noise generated by fluorescent scattering from atoms outside the tomographic section in a three-dimensional sample---is proposed, and is modeled using the simulation developed in the current article.

In fact the fluorescent imaging of atoms undergoing in-lattice polarization gradient cooling is a measurement technique which has the potential to be used in a wider class of experiments than just the resolved-atom experiments described above. The ability to extract a very strong signal from each atom in a dilute sample may be useful in a variety of scenarios; for example, to make a time-of-flight (column-density) measurement on a very low atom-number sample which is unobservable using absorption imaging due to its dilution. Unlike for resolved-atom measurements on dense samples, very high spatial resolution is likely not needed, so simpler apparatus may be used.

The structure of this article is as follows. A conceptual outline of polarization gradient cooling for an atom in an optical lattice is given in Section \ref{SDcoolconf}, with emphasis on how the lattice affects the cooling process. The system is described theoretically in Section \ref{MCsimul}, and the dynamical equations are rendered suitable for computation by the use of a semi-classical Monte Carlo wavefunction method. In Section \ref{ext_MCSect} the hybrid Monte Carlo-Master Equation (HMCME) method is developed in order to predict the atomic hopping rate between wells. In Section \ref{FricSect} the frictional force felt by atoms undergoing polarization gradient cooling in a far-detuned optical lattice is calculated and compared to the case of untrapped atoms. A calculation of the site hopping rate is presented for a specific set of parameters in Section \ref{ExampDyn}. The dependence of the site confinement time on the cooling configuration and system parameters is investigated in Section \ref{lifeHop}, and conclusions are drawn. 

\section{Polarization gradient cooling of atoms trapped in a lattice potential}
\label{SDcoolconf}
\subsection{Polarization gradient cooling without the additional potential}

Polarization gradient cooling exploits the polarization gradients of light fields to cool atoms. We briefly review two specific types of polarization gradient cooling which were analyzed in a seminal paper by Dalibard and Cohen-Tannoudji \cite{dalibard_laser_1989}: the one-dimensional orthogonal linear (lin$\,{\perp}\,$lin) and orthogonal sigma ($\sigma^{+}\sigma^{-}$) configurations for two equal intensity counter-propagating laser beams.

\begin{figure*}
\subfloat[Without pinning lattice]{\hspace*{1.2cm}\label{nextCool1}\includegraphics[scale=0.64]{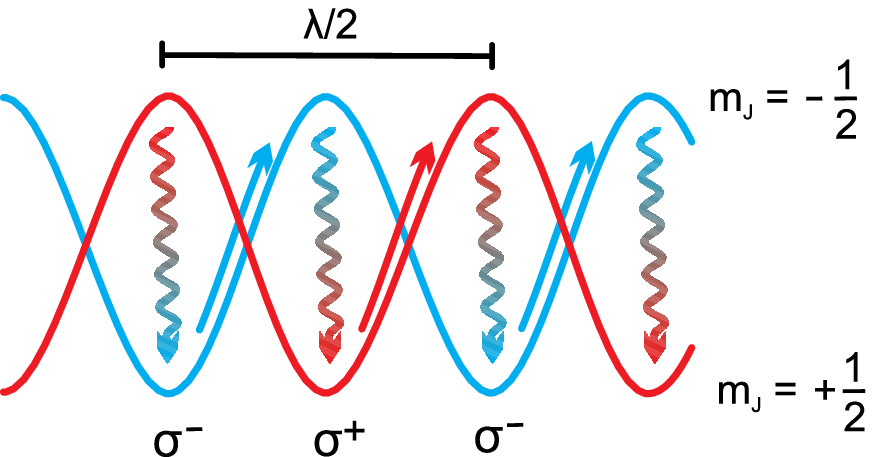}}\\
\subfloat[In pinning lattice, minima coincide ($\phi\,$=$\,0$) ]{\label{nextCool2}\includegraphics[scale=0.56]{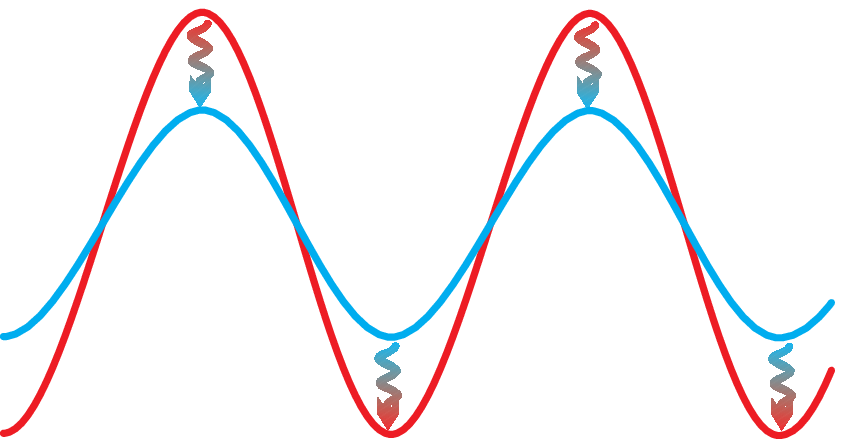}}\\
\subfloat[In pinning lattice, minima displaced ($\phi\,$=$\,\pi/4$)]{\label{nextCool3}\includegraphics[scale=0.56]{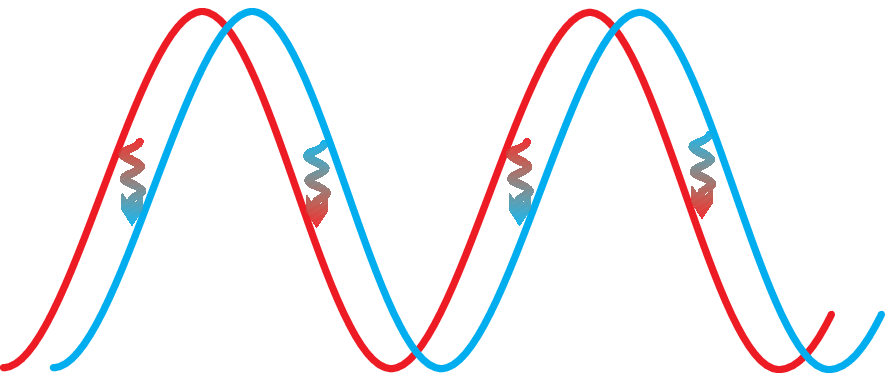}}
\caption{(Color online) Polarization gradient cooling in the 1D lin$\,{\perp}\,$lin configuration for an atom with $J_{g}\,$=$\,1/2$. (a) Without pinning lattcice. The atom loses kinetic energy as it climbs the potential; near the top of the potential, it is optically pumped into the lower potential, losing energy over the complete cycle. (b) Confined by a pinning lattice (with spatial period $\lambda_{l}/2$) with minima coinciding with the state-dependent lattice formed by the cooling light ($\phi\,$=$\,0$). The cooling is inefficient for atoms which have less than half of the lattice binding energy. (c) Confined by a pinning lattice (with spatial period $\lambda_{l}/2$) with offset minima ($\phi\,$=$\,\pi/4$). The low-energy curve crossing enables deeply-bound atoms to be cooled efficiently.}
\label{coolTDiag}
\end{figure*}

For the lin$\,{\perp}\,$lin configuration, the positive component of the electric field is
\begin{equation}\label{E1}
{\bm E^{+}}\propto \cos \left(kz + \phi \right) \, {\bm e_{-1}} -i \sin \left(kz + \phi \right) \, {\bm e_{+1}}
\end{equation}
where ${\bm e_{\pm 1}}\,$$=\,$$\mp \left(1/\sqrt{2}\right)\left(\hat{\bm x}\pm i\hat{\bm y}\right) $. The light potential consists of alternating standing waves of $\sigma^{+}$ and $\sigma^{-}$ light. For a ground-state atom with a spin of $J_{g}\,$=$\,1/2$, the $\sigma^{+}$ standing wave couples more strongly to the $m_{J}\,$=$\,+1/2$ state, and vice versa; the ground-state light shifts therefore oscillate alternately according to position (see Fig.\ \ref{nextCool1}). Furthermore, the $\sigma^{+}$ ($\sigma^{-}$) light pumps atoms towards the $m_{J}\,$=$\,+1/2$ ($m_{J}\,$=$\,-1/2$) state, which for negative laser frequency detuning, lies below the original state. The motion of the atom is therefore described by a loss in translational energy as the atom climbs the state-dependent lattice potential, followed by optical pumping from a peak to a trough in the lattice potential accompanied by a much smaller recoil momentum exchange (see Fig.\ \ref{nextCool1}). This `Sisyphus' process is dissipative.

For the $\sigma^{+}\sigma^{-}$ configuration, the positive component of the electric field is
\begin{equation}\label{E2}
{\bm E^{+}}\propto \cos \left(kz + \phi \right) \, \hat{\bm x} - \sin \left(kz + \phi \right) \, \hat{\bm y} \; .
\end{equation}
This polarization is always linear, but orientated at an angle which is proportional to $z$; it is convenient to think of a spatially varying atomic orientation axis parallel to the electric field, so the light is $\pi$ polarized everywhere. While the lights shifts are constant, they are not equal for ground state atoms with spin $J_{g}\,$$\geq$$\,1$; the Clebsch-Gordon coefficients are such that, with the laser tuned below resonance, the energy offset of the states increases monotonically with $|m_{J}|$. Optical pumping preferentially transfers population from the high $|m_{J}|$ (higher energy) states to the low $|m_{J}|$ (lower energy) states. As the atom moves, the new basis is different to that of the original basis. If the atom starts in a low $|m_{J}|$ state in the original basis, the atom has a higher proportion of its density matrix in the higher $|m_{J}|$ states in the new basis; these lie higher in energy, and the atom feels a net force opposing its motion. The optical pumping preferentially returns the atom into a low $|m_{J}|$ state at the new position, so the net force on an atomic trajectory is again dissipative.

\subsection{One dimensional cooling in a lattice using orthogonal linear polarizations}
\label{orthlin}

For an atom confined to a region smaller than a wavelength the nature of the polarization gradient cooling processes are modified. In the cases under consideration the atom is confined by an additional optical lattice potential, which shall be called the pinning lattice; the pinning lattice has an optical frequency far away from atomic resonances, so the mixing of the ground and excited states induced by the pinning lattice potential is negligible compared to the mixing induced by the cooling beams, and consequently scattering from the lattice potential is negligible. It is assumed that the pinning lattice potential is state-independent (or nearly so) for ground-state atoms, so the frequency detuning of the lattice is much greater than the width of the hyperfine structure of the excited state. 

In this section we consider a pinning lattice formed by counter-propagating laser beams; the lattice spatial period is $\lambda_{l}/2$, where $\lambda_{l}$ is the wavelength of the lattice laser beams. The counter-propagating lin$\,{\perp}\,$lin cooling light induces an additional state-dependent lattice potential with spatial period $\lambda_{c}/2$, where $\lambda_{c}$ is the wavelength of the cooling light (in this article the cooling and lattice laser wavelengths will be similar but not equal, $\lambda_{l}\,$$\approx$$\,\lambda_{c}\,$). The pinning lattice potential is characterized by its depth and phase with respect to the cooling light (this phase will be a function of position if the lattice periods are not equal). It is assumed that the pinning lattice is substantially deeper than the potential induced by the cooling light. The pinning lattice may be in phase (Fig.\ \ref{nextCool2}) or out of phase (Fig.\ \ref{nextCool3}) with the lattice potential produced by the cooling light; these two situations have very different cooling characteristics.

For the case of $\phi\,$=$\,\pi/4$, there are equal intensities of $\sigma^{+}$ and $\sigma^{-}$ light at the minima of the combined potential, with equal and opposite intensity gradients, so the potential minima for $m_{J}\,$=$\,\pm 1/2$ are equal in energy, but offset a small distance from each other (Fig.\ \ref{nextCool3}). For laser light with a negative frequency detuning with respect to the atomic resonance, atoms are optically pumped from the upper potential to the lower potential. Therefore for $\phi\,$=$\,\pi/4$ atoms are polarization gradient cooled efficiently at low energies as the optical pumping direction reverses near the potential minima; Sisyphus-like cycles can be performed even by deeply bound atoms.

For the case of $\phi\,$=$\,0$, the optical pumping changes sign half way up the pinning lattice potential (Fig.\ \ref{nextCool2}) so Sisyphus effects do not efficiently cool atoms with an energy below this point. Cooling can occur for atoms with energy greater than this point; however calculations will show (Figure \ref{coolTV}) that this is less efficient than for the $\phi\,$=$\,\pi/4$ case. The equilibrium temperature of atoms undergoing lin$\,{\perp}\,$lin polarization gradient cooling with $\phi\,$=$\,0$ will be much higher than for those with $\phi\,$=$\,\pi/4$.

\label{freqChanDir}
Another difference in the polarization gradient cooling of lattice-trapped atoms is that the additional pinning lattice potential rapidly changes the velocity and position of the atom. If the intensity of the cooling light is such that the optical pumping time is greater than half of a lattice oscillation period, the atom cannot undergo a full optical pumping cycle before the relative energies of the states are reversed by the motion of the atom (see Fig.\ \ref{nextCool3}), so the cooling efficiency will be decreased if the cooling beams are not intense enough, and the steady-state temperature of the atoms is expected to increase at sufficiently low cooling intensity (see Figs.\ \ref{VarInten1D} and \ref{VarInten3D}). In comparison, consider that without the lattice the atom encounters similar light fields at a rate of approximately $2|v|/\lambda_{c}$, while the lattice-bound atom encounters similar light fields at a rate of approximately $\omega_{osc} /\pi$. Thus the intensity below which cooling becomes degraded is affected by the frequency of the lattice oscillations; for deep enough lattices (i.e.\ fast enough oscillations) this intensity may be expected to be higher than is the case with no lattice present.

A further point is that the polarization gradient cooling process relies on inducing a differential dipole potential on the sublevels in order to generate the cooling force. If the pinning lattice has a sublevel-dependent component then this modifies the cooling force, and can potentially disrupt the cooling process altogether.

\subsection{One dimensional cooling in a lattice using orthogonal circular polarizations}

The $\sigma^{+}\sigma^{-}$ configuration in one dimension, unlike the lin$\,{\perp}\,$lin configuration, does not depend on the relative phase of the lattices as the light shift from the molasses is the same everywhere. A change in phase of the molasses is identical to a rotation about the axis of the laser beams; by a symmetry argument, it can be seen that the dynamics are unchanged in the direction parallel to the laser beams. (There is a redistribution of the components of spontaneously emitted radiation in the orthogonal directions upon such a phase change, but the effect of this on the cooling dynamics is minor). As with the lin$\,{\perp}\,$lin configuration, the optical pumping should occur on a time scale shorter than half the period of oscillation for efficient cooling. 

\subsection{Three-dimensional polarization gradient cooling of lattice-confined atoms}
\label{cool3dinit}

Polarization gradient cooling can also take place if atoms are exposed to 3 sets of counter-propagating beams. The atoms are exposed to both polarization and intensity gradients in 3D molasses light; the polarization gradients cool the atoms by an admixture of the lin$\,{\perp}\,$lin and $\sigma^{+}\sigma^{-}$ mechanisms \cite{salomon_laser_1990}. 

Tightly confined atoms interact with the cooling light within a volume smaller than a cubic wavelength, so the net intensity and polarization that an atom experiences depends on all five of the relative phases of the molasses beams. As a consequence the cooling efficiency along each of the three orthogonal directions at each pinning lattice site depends on the lattice site position together with these relative phases; while some atoms will be cooled efficiently, others will be cooled only weakly. Furthermore atoms at different positions within the molasses light field will fluoresce at different rates; this can pose a problem when correlating fluorescent signal to atom numbers.

This situation is unacceptable when the atoms are required, to a high probability, to be localized to each site for a long time. This article discusses two ways to overcome this problem. The first way would be to turn each set of counter-propagating one-dimensional $\sigma^{+}\sigma^{-}$ beams on sequentially, cooling the atoms along each direction in turn. The second way would be to introduce a small frequency difference between each pair of beams; the phases of the cooling light at any point, and so the molasses character, will then change continuously. This is the method used in recent experiments \cite{bakr_quantum_2009,bakr_probing_2010,sherson_single_2010}. The two methods are analyzed in Sections \ref{ExampDyn} and \ref{lifeHop}. 

\section{Monte Carlo simulation of in-lattice polarization gradient cooling}\label{MCsimul}
This section outlines the Monte Carlo simulation of the polarization gradient cooling of atoms confined in an additional optical potential. 

The calculations presented in this article are performed for a single isolated atom, greatly simplifying the theoretical description of the system. Reabsorption of scattered radiation is possible amongst an extended atomic sample; the influence of rescattering for a particular sample may be estimated by the ratio of the rescattered light to the incident light intensity at an atom. However, the atom number and density will be low in the situations under consideration, so this ratio will typically be much less than unity---the object of the measurement is to determine the position of each atom in a sample, and this will not be possible if there are too many atoms, or they are too closely packed. 

\subsection{Simplification of the simulated system}

It is assumed that the additional confining optical potential has a frequency far away from the atomic resonance, so the mixing of the ground and excited states induced by the potential is negligible compared to the mixing induced by the cooling beams; consequently fluorescent scatter from the optical potential is negligible. As a further consequence mixing between internal states induced by the optical potential is ignored---of course, this mixing must be present to generate the potential, but in the far-detuned regime this mixing is much smaller than the mixing due to the cooling beams, and is ignored. 

In this article it is assumed that the additional optical potential is a three-dimensional optical lattice (the pinning lattice), with three pairs of beams frequency detuned from each other by many megahertz, so the effective potential is the sum of the one-dimensional potentials. The potential generated by the optical lattice is therefore
\begin{equation}
{\bm V}({{\bm r}})={\bm V}_{\bm 0}\left (\sin^{2}{kx}+\sin^{2}{ky}+\sin^{2}{kz} \right) \; ,
\end{equation}
where ${\bm V}_{\bm 0}$ is a matrix dependent on the hyperfine structure of the transition; in general it is state-dependent, and includes non-zero off-diagonal ground-ground or excited-excited elements dependent on the optical polarization. 

The majority of alkali species have two ground hyperfine levels. Cooling takes place on a closed transition, i.e.\ $|g,I\,$$+\,$$1/2\rangle\,$$\rightarrow\,$$|e,I\,$$+\,$$3/2\rangle$, where $I$ is the nuclear spin; however, the atoms also have a small probability of being excited to the $|e,I\,$$+\,$$1/2\rangle$ state, which can decay to the lower hyperfine state $|g,I\,$$-\,$$1/2\rangle$. The atoms will eventually be optically pumped to this dark state. As usual, this behavior is to be prevented by the presence of `repumping' light, i.e.\ light resonant with the $|g,I\,$$-\,$$1/2\rangle\,$$\rightarrow\,$$|e,I\,$$+\,$$1/2\rangle$ transition. It is assumed that this repumping light is of sufficient power that the total residence time in the $|g,I\,$$-\,$$1/2\rangle$ state is only a few times the natural lifetime of the atom in the exited states; consequently there is an extremely small population of atoms in the lower hyperfine state at any one time. This means that the time spent in the lower hyperfine level is much less than the period of vibration in the lattice, so negligible heating of the atom will occur due to differences in the optical potentials of the two ground hyperfine states. The repumping light is off-resonance by some gigahertz from the cooling transition $|g,I\,$$+\,$$1/2\rangle\,$$\rightarrow\,$$|e,I\,$$+\,$$3/2\rangle$, and as such will only have a very small effect on the dynamics of the atoms within those states, which is ignored in the subsequent analysis. The population of the $|e,I\,$$+\,$$1/2\rangle$ excited state is also ignored.

\subsection{Unitary quantum dynamics}

The Hamiltonian describing the unitary evolution of the atomic state in the electric field of the cooling light in the rotating wave approximation is
\begin{equation} \label{main_ham}
{\bm H}=-\frac{\hbar^{2}{\bm I}}{2m}\nabla^{2}+{\bm V_{0}}\left (\sin^{2}{kx}+\sin^{2}{ky}+\sin^{2}{kz} \right )-\hbar \Delta{\bm P}+\frac{\hbar \Omega}{2}\sum_{\varepsilon=-1,0,1}\left (E_{\varepsilon}({\bm r}){\bm D^{\bm +}_{\bm \varepsilon}}+E^{*}_{\varepsilon}({\bm r}){\bm D^{\bm -}_{\bm \varepsilon}} \right) \; .
\end{equation}
The state vector contains $4I\,$$+\,$$6$ components describing the $|g,I\,$$+\,$$1/2\rangle$ and $|e,I\,$$+\,$$3/2\rangle$ states. The matrix ${\bm V_{\bm 0}}$ is the state-dependent potential, and ${\bm I}$ the identity matrix. The matrices ${\bm D^{+}_{\varepsilon}}$ (${\bm D^{-}_{\varepsilon}}$) are the dimensionless raising (lowering) matrices with values given by the Clebsch-Gordon coefficients. These matrices are normalized so that
\begin{equation}
{\bm P}=\sum_{\varepsilon=-1,0,1} {\bm D^{\bm +}_{\bm \varepsilon}}{\bm D^{\bm -}_{\bm \varepsilon}} \; ,
\end{equation}
in which ${\bm P}$ is the excited state projection operator. The relative normalization of the local electric field vector $E_{\varepsilon}({\bm r})$ and the Raman frequency parameter $\Omega$ is arbitrary. The frequency detuning from resonance of the cooling light is $\Delta$.

The evolution of the density matrix ${\bm \rho}$ due to unitary processes obeys the von Neumann equation
\begin{equation} \label{main_ham_evol}
i \hbar \frac{\partial}{\partial t} {\bm \rho}= \left [ {\bm H}, {\bm \rho}\right ]\; .
\end{equation}

\subsection{Relaxation mechanisms of the internal states}

The effects of the relaxation processes are determined from a trace over the `environmental' degrees of freedom of the corresponding unitary processes. To account for recoil from spontaneous emission processes, recoil momentum terms are added to the standard result \cite{cohen-tannoudji_atomic_1992} in a sum over all transitions and polarizations
\begin{equation} \label{sp_dyn_rec}
\left (\frac{\partial}{\partial t} {\bm \rho} \right)_{sp}= \Gamma \left ( -\frac{1}{2}\left ( {\bm P}{\bm \rho} +{\bm \rho}{\bm P} \right) +\sum_{\varepsilon,\varepsilon',\sigma} \int d^{2}{\bm \kappa} \: e^{i k_{R}{\bm \kappa}.{\bm r'}} \: {\bm D^{\bm -}_{\bm \varepsilon}}{\bm \rho}{\bm D^{\bm +}_{\bm \varepsilon'}} \: e^{-i k_{R}{\bm \kappa}.{\bm r}} \: f^{\sigma}_{\varepsilon \varepsilon'}({\bm \kappa}) \right) \; .
\end{equation}
Here ${\bm \kappa}$ is a unit vector centered on the origin, and the parameter $f^{\sigma}_{\varepsilon \varepsilon'}({\bm \kappa})$ describes the angular distribution of the spontaneous emission of photons with polarization $\sigma$ for a particular combination of raising and lowering operators $\varepsilon$ and $\varepsilon'$. The magnitude of the laser wavevector is denoted by $k_{R}$, and $\Gamma$ is the angular frequency width of the excited state. The parameters $f^{\sigma}_{\varepsilon \varepsilon'}({\bm \kappa})$ are derived in Appendix \ref{dir_sp}.

\subsection{Explicit retention of the excited states}
\label{retentUpper}

In many analyses of polarization gradient cooling (e.g.\ \cite{dalibard_laser_1989,grynberg_cold_2001}) the excited states are adiabatically eliminated. This is appropriate for investigations that look to find the lowest temperature of atoms in optical molasses, as this usually occurs at low intensity; at low intensities, the population of the excited states is small, and adiabatic elimination is a good approximation. However, a high scattering rate is preferable for efficient signal extraction during the measurement process; furthermore, as discussed in Section \ref{freqChanDir}, the lowest temperatures of confined atoms tend to occur at higher intensity than for untrapped atoms. Therefore, the excited state populations are retained in the analysis in order to accurately simulate the situations of interest. 

\subsection{The Wigner transformation}

In anticipation of taking the semi-classical approximation, the quantum dynamics of Equations \ref{main_ham} and \ref{sp_dyn_rec} are expressed in terms of the Wigner function
\begin{equation} \label{wiga}
{\bm W}({\bm r},{\bm p},t)=\frac{1}{h^3} \int d^{3}{\bm u}\left\langle {\bm
r}+\frac{\bm u}{2} \right| {\bm \rho} \left| {\bm r}-\frac{\bm u}{2}
\right\rangle e^{- i {\bm p}\cdot {\bm u}/\hbar}\; .
\end{equation}
It is shown in Appendix \ref{spont_append} that the complete dynamical equation in terms of the Wigner function is
\begin{eqnarray} \label{wig_comp}
&&\left( \frac{\partial}{\partial t} +\frac{{\bm p}}{m} \cdot \nabla \right) {\bm W}({\bm r},{\bm p},t)=\\&&\quad\quad\frac{2\pi i}{h^{4}} \int d^{3}{\bm s} 
e^{i{\bm p}\cdot{\bm s}/\hbar} \left ( \left\langle {\bm
r}+\frac{\bm s}{2} \right| {\bm \rho} \left| {\bm r}-\frac{\bm s}{2}
\right\rangle {\bm V}\left({\bm r-\frac{\bm s}{2}}\right) - {\bm V}\left({\bm r+\frac{\bm s}{2}}\right) \left\langle {\bm
r}+\frac{\bm s}{2} \right| {\bm \rho} \left| {\bm r}-\frac{\bm s}{2}
\right\rangle \right ) \nonumber\\
&&\quad\quad-\frac{\Gamma}{2}\left ( {\bm P}{\bm W} +{\bm W}{\bm P} \right) +\Gamma\sum_{\varepsilon,\varepsilon',\sigma} \int d^{2}{\bm \kappa} {\bm D^{\bm -}_{\bm \varepsilon}}{\bm W}({\bm r},{\bm p}-\hbar k_{R} {\bm \kappa},t){\bm D^{\bm +}_{\bm \varepsilon'} } f^{\sigma}_{\varepsilon \varepsilon'}({\bm \kappa}) \nonumber \; ,
\end{eqnarray}
where
\begin{equation} 
{\bm V({\bm r})}={\bm H}+\frac{\hbar^{2}{\bm I}}{2m}\nabla^{2} \;.
\end{equation}
The Hamiltonian ${\bm H}$ is given by Equation \ref{main_ham}.

\subsection{The semi-classical approximation}\label{semiclass_approx}

A full quantum treatment of the problem is computationally infeasible; instead, the semi-classical approximation will be used. In the semi-classical approximation the external degrees of freedom of the atom are treated classically, while the internal degrees of freedom are treated quantum mechanically. The semi-classical approximation is used in most theoretical treatments of polarization gradient cooling \cite{dalibard_laser_1989,cohen-tannoudji_atomic_1992,petsas_semiclassical_1999,grynberg_cold_2001,jonsell_nonadiabatic_2006}. The approximation can be used in situations in which the coherence length of the atomic ensemble is much less than a wavelength; it relies on the dephasing influence of spontaneous emission to in effect `localize' the atoms. The accuracy of this approximation as used in the situation under consideration is discussed in Section \ref{discussSimDyn}.

Using the semi-classical approximation, the equations of motion are expanded in the small parameter $\epsilon_{1}=\hbar k_{R}/\Delta p$, with higher orders in the expansion discarded \cite{cohen-tannoudji_atomic_1992}; here $\Delta p$ is the momentum width of the Wigner function. For example, the expansion of the potential term on the right hand side of Equation \ref{wig_comp} is
\begin{equation} \label{wig_exp}
{\bm V}({\bm r}+{\bm s'})= {\bm V}({\bm r}) + {\bm s'}\cdot \nabla{\bm V}({\bm r})+ \frac{1}{2} \left ( {\bm s'}\cdot \nabla\right )^{2}{\bm V}({\bm r})+\ldots \; .
\end{equation}
This quantity predominately contributes to the integral in the region $|{\bm s'}|<\hbar/\Delta p$. The derivative of the potential $V$ has approximate magnitude $k_{R}V$, so the series expansion is seen to be in the small parameter $\epsilon_{1}$; it is terminated at appropriate order, which is chosen in this analysis to be the second order. 

This approximation leads to the equation (Appendix \ref{ham_append})
\begin{eqnarray} \label{wig_app}
\left( \frac{\partial}{\partial t} +\frac{{\bm p}}{m} \cdot \nabla \right) {\bm W}&=& \frac{i}{\hbar} \left [ {\bm W}, {\bm V} \right] +\frac{1}{2} \sum_{i} \left \{ \frac{\partial {\bm W}}{\partial p_{i}}, \frac{\partial {\bm V}}{\partial r_{i}} \right \} \\ &-&\frac{i \hbar}{8} \sum_{ij} \left [ \frac{\partial^{2} {\bm W}}{\partial p_{i}\partial p_{j}}, \frac{\partial^{2} {\bm V}}{\partial r_{i}\partial r_{j}} \right ] + \Gamma \sum_{\varepsilon}{\bm D^{\bm -}_{\bm \varepsilon}} {\bm W}{\bm D^{\bm +}_{\bm \varepsilon} } \nonumber \\ \nonumber &-&\frac{\Gamma}{2}\left ( {\bm P}{\bm W} +{\bm W}{\bm P} \right) +\frac{\Gamma \hbar^{2}k_{R}^{2}}{2}\sum_{\varepsilon,\varepsilon',i,j} \eta_{\varepsilon \varepsilon' i j}{\bm D^{\bm -}_{\bm \varepsilon}}\frac{\partial^{2} {\bm W}}{\partial p_{i}\partial p_{j}}{\bm D^{\bm +}_{\bm \varepsilon'} } \; .
\end{eqnarray}
In this equation, the curly brackets are the anti-commutator, and the tensor $\eta_{\varepsilon \varepsilon' i j}$ is given in Appendix \ref{dir_sp}. 

\subsection{Conversion to Langevin form}

The semiclassical evolution equation as it stands (Equation \ref{wig_app}) still requires substantial computational power to simulate. Instead, the calculation will be restricted to a single trajectory; the distribution of atomic properties is then found by the sum over these trajectories. This approach is stochastic i.e.\ Monte Carlo in nature. The equations of motion for a specific trajectory in position and momentum space are found by substituting a semiclassical trial solution
\begin{equation} \label{delta_app}
{\bm W}( {\bf r},{\bf p},t)={\bm w}(t)\;\delta^{(3)}({\bf r}-\tilde{\bm r})\;\delta^{(3)}({\bf p}-\tilde{\bm p}) \; ,
\end{equation}
into Equation \ref{wig_app}. This solution is valid in the limit in which both parameters $\epsilon_{1}=\hbar k_{R}/\Delta p$ and $\epsilon_{2}=k_{R}\Delta p /m\Gamma$ are small, as discussed in Reference \cite{cohen-tannoudji_atomic_1992} and in the previous section. The parameter $\epsilon_{2}\approx\epsilon_{1} T/T_{D}$, where $T_{D}$ is the Doppler temperature, is much smaller than one in the situations considered in this article.

Integration over external co-ordinates gives the equation of evolution for the internal co-ordinates
\begin{equation} \label{internal_app}
\frac{\partial}{\partial t} {\bm w} = \frac{i}{\hbar}\left [{\bm w},{\bm V}(\tilde{\bm r}) \right]+\Gamma \left ( -\frac{1}{2}\left ( {\bm P}{\bm w} +{\bm w}{\bm P} \right) +\sum_{\varepsilon} {\bm D^{\bm -}_{\bm \varepsilon}}{\bm \rho}{\bm D^{\bm +}_{\bm \varepsilon}} \right) \; .
\end{equation}
The equation of evolution of the external co-ordinates, e.g.\ $\tilde{\bm r}=\langle {\bm r} \rangle$, are found to be
\begin{eqnarray} \label{motion}
\frac{\partial \tilde{r}_{i}}{\partial t}&=&\frac{ \tilde{p}_{i}}{m}\\
\frac{\partial \tilde{p}_{i}}{\partial t}&=&f_{i}\;=\;-\textnormal{Tr}_{I} \left( {\bm w}\frac{\partial {\bm V}(\tilde{\bm r})}{\partial r_{i}}\right) \label{motion2}\\
\frac{\partial}{\partial t} \left \langle (r_{i}-\tilde{r}_{i}) (r_{j}-\tilde{r}_{j}) \right \rangle&=&0\\
\frac{\partial}{\partial t} \left \langle (p_{i}- \tilde{p}_{i})(p_{j}- \tilde{p}_{j}) \right \rangle
&=&2D_{ij}\;=\;\hbar^{2}k^{2}_{R}\Gamma\frac{1+\delta_{ij}}{2} \sum_{\varepsilon,\varepsilon'} \eta_{\varepsilon \varepsilon' i j}
\textnormal{Tr}_{I} \left( {\bm D^{\bm -}_{\bm \varepsilon}}{\bm w}{\bm D^{\bm +}_{\bm \varepsilon'}}\right) \; . \label{motion_end}
\end{eqnarray}

The equation containing $D_{ij}$ describes the diffusion of the atom due to the atomic recoil. The diffusion term is a fluctuating Langevin force with zero mean, and is incorporated into the motion of a single trajectory in the It\={o}-Langevin equation
\begin{equation} \label{ito_lang}
dp_{i}=f_{i}dt+\sum_{k}\sqrt{2d_{ik}}dW_{k}
\end{equation}
in which $dW_{j}$ are independent, zero mean, Gaussian-distributed stochastic increments with variance $dt$. The quantities $d_{ik}$ are the components of the $k$th eigenvector of $D_{ij}$, which is normalized according to its eigenvalue.

\subsection{Unraveling the Optical Bloch equations}
\label{unrav}

Equations \ref{motion} to \ref{ito_lang} describe the motion of a single atom in the semi-classical approximation. However, the internal dynamics as described in \ref{internal_app} are still ensemble-averaged, and are not appropriate to describe a single trajectory. The optical Bloch equation (Equation \ref{internal_app}) must therefore be unraveled into the stochastic evolution of a single wave function. The unraveling is chosen to be that of the quantum Monte Carlo wave function method (QMCW) \cite{molmer_monte_1993}. The underlying principle of the approach is to evolve the wave function using a non-Hermitian quasi-Hamiltonian with the addition of randomly-occurring discrete quantum jump processes. 

In the QMCW approach the following procedure is followed for each time step. Firstly, a random number 0$\,<\,$$r$$\,<\,$1 chosen from a flat distribution and is compared to the quantity $j=1-\Gamma \delta t \langle \psi | {\bm P} | \psi \rangle$. If $r$$\,<\,$$j$,
\begin{equation} \label{evol}
| \psi (t+\delta t) \rangle = | \psi (t) \rangle - \frac{i}{\hbar} \delta t \left({\bm V}-\frac{i\hbar \Gamma}{2}{\bm P}\right) | \psi (t) \rangle.
\end{equation}
If the $r$$\,>\,$$j$, the atom undergoes a quantum jump (photon emission); photon polarization is chosen randomly according to the weights 
\begin{equation} \label{evol_weights}
p_{\varepsilon}=\frac{\langle \psi | {\bm D^{\bm +}_{\bm \varepsilon}}{\bm D^{\bm -}_{\bm \varepsilon}} | \psi \rangle}{\langle \psi | {\bm P} | \psi \rangle} \; ,
\end{equation}
and the new wave function is given by 
\begin{equation} \label{proj_out}
 | \psi (t+\delta t) \rangle = {\bm D^{\bm -}_{\bm \varepsilon}} | \psi (t) \rangle.
\end{equation}
The new wave function requires normalization in either case. It can be shown that the ensemble of possible new wavefunctions satisfies the optical Bloch equation (Equation \ref{internal_app}).

\subsection{Computational methods}\label{compMeth}

The set of equations \ref{motion} to \ref{proj_out} specify the stochastic evolution of a single semiclassical trajectory. An explicit third-order Runge-Kutta method is used to propagate the classical position and momentum of the atom. The coefficients for the method are chosen so that the intermediate evaluation times are $h/3$ and $2h/3$, in which $h$ is the time step for the external dynamical evolution. The internal components of the state vector are advanced at a constant time step $h/3$, and are propagated by the Cayley (split) form of the evolution operator. The new wave function is found efficiently by Gaussian elimination.

The method outlined in this section was tested against simple analytic models and by comparison against previously published work looking at polarization gradient cooling in one-dimensional dissipative optical lattices \cite{castin_quantization_1991,petsas_semiclassical_1999,jonsell_nonadiabatic_2006}. The results agree very well for intermediate saturation parameters. At high and low saturation parameters, the results differ somewhat; the retention of the excited states means the method presented here is more accurate for high saturation parameters ($s\,$$>$$\,$0.1), while the use of adiabatic elimination in the previous work enabled better statistics for lower saturation parameters ($s\,$$<$$\,$0.02). The Monte Carlo simulation was also tested against previous data for polarization gradient cooling of rubidium atoms in three-dimensional molasses \cite{javanainen_polarization_1994}; for intermediate saturation parameters the predicted temperatures agreed to within 10\%, which is around the error quoted for the previous work.

\section{Hybrid Monte Carlo-Master Equation analysis}
\label{ext_MCSect}
\subsection{Comment on nature of the simulation}
\label{commNat}

The problem under consideration differs from previous analyses of polarization gradient cooling in a number of respects. The difference which poses the greatest computational challenge is the ratio of the time scales in the problem---between the phenomena of interest, the jumping of atoms between wells, which we would like to take place over tens of seconds or greater; and the smallest time scale relevant to the problem, the period of the beat frequency between the laser and atomic resonance, which will be in the range of 10$\,$ns to 100$\,$ns. This means that, on average, one event of interest will occur every $10^{9}$ to $10^{12}$ time steps. Clearly a straightforward Monte Carlo simulation of these phenomena will be prohibitively slow.

Adiabatic elimination is often used in simulations of polarization gradient cooling in order to increase the size of the smallest time step (and decrease the size of the state vector). However, this technique can only be used when the saturation parameter is much less than 1, which is not the case for the situations under consideration in this article (see Sect.\ \ref{retentUpper}). In any case, the gain from this technique would be, at most, one or two orders of magnitude in the time scale ratio, i.e.\ interesting events would occur every $10^{7}$ to $10^{10}$ time steps; such calculations would still be very computationally intensive.

\subsection{An extended Monte Carlo analysis for rare events}
\label{ext_MC}

In order to make the simulation tractable a method has been developed to extend the Monte Carlo simulation in order to study rare events. This technique bears similarity to previously developed techniques which are collectively known as the `splitting' methods of Monte Carlo simulation \cite{lecuyer_rare_2007,villen-altamirano_analysis_2002,garvels_importance_2002,glasserman_multilevel_1999}. 

The basic idea is that a non-Markovian system can look Markovian when its dynamics are averaged over a sufficiently long time interval. An atom undergoing fluorescent scattering is clearly non-Markovian at timescales shorter than the scattering period; the deterministic state-dependent dynamics of the Bloch vector dominates, with stochastic dynamics (environmental coupling) playing only a minor role. However, when viewed over sufficiently long time intervals, it is known (from theoretical and experimental work) that quantities such as temperature can be assigned to an atom undergoing polarization gradient cooling; for such an assignment to make sense these quantities must be independent of the state of the atom at any one time, so implicitly it is assumed that the atomic dynamics, on long timescales, is described by stochastic Markovian dynamics. 

The hybrid Monte Carlo-Master Equation (HMCME) method presented in this section uses this idea to extend Monte Carlo simulation so that it is capable of predicting the frequency of rare events. The method is approximate; it uses the assumption that the system dynamics are Markovian when viewed over appropriately long time intervals.

\subsection{Classification according to energy}
\label{measEv}

Firstly, a quantity should be found that is representative of the aspect of the system which is to be investigated; it should be a scalar time-dependent quantity which is a function of the state vector. It is chosen with two properties in mind; that it varies slowly and smoothly with time, and that the rare events of interest occur at values well separated from the usual values it takes. For confined atoms undergoing polarization gradient cooling, an energy-like quantity is appropriate; the atoms usually have energies well below that required to hop to a neighboring site. In the context of the subsequent analysis, the approximation will be used that the system performs Markovian dynamics---a random walk---along the energy axis. The Markovian approximation implies that the system is ergodic over long timescales.

It is desired to label the system according to a discrete energy parameter, so a set of `points' are defined at certain energies, $\{E_{1}, E_{2}, \mathellipsis \}$. As the time evolution is simulated using a Monte Carlo method, the system will encounter these points repeatedly. (If the state vector of the system undergoes discontinuous jumps, the time and state vector of the system as it crosses a point may need to be extrapolated or approximated). The state of the system at any time will be classified by the last point encountered by the system. Each time the system encounters one of these points the state vector (the complete description of the system) is recorded; these records will be called `start vectors'.

\subsection{Finding representative start vectors}

A representative set of start vectors are required for each point, i.e.\ a set which fairly samples the true population of start vectors for that point. However, the run time of the Monte Carlo simulation should also be minimized. As the system encounters points at higher energies only very rarely, it would be prohibitive to take data for these points by means of direct Monte Carlo simulation. These two considerations need to be balanced.

As a first stage, the simulation is run from some arbitrary starting point until it settles down into a steady state condition. From this time on the state vector is recorded every time the system encounters a point $E_{i}$. A representative set of start vectors will be built up for points at low energy.

\subsection{Focusing the analysis on higher energies}
\label{sevDau}

This coverage is now extended to points lying at higher energy. To do this a weight $w_{ni}$ is assigned to each start vector; this is an estimate of how representative that start vector is of the general population of start vectors at all points. Initially an equal weight is assigned to each of the extant start vectors; the weights are normalized. The total weight of the start vectors at a particular point $W_{i}$ is related, but not equal, to the probability of finding the steady state system in the environs of point $i$.

The nature of the simulation is now changed so that new start vectors (`daughter' start vectors) are to be found from existing start vectors (`mother' start vectors). To do this, a start vector is picked at random from all the possible start vectors according to its weight, and is evolved in time until a new point is reached. As it is assumed that the system is ergodic, the `mother' start vector at the old point can be replaced by the new `daughter' start vector at the new point (by assigning the mother start vector's original weight to the daughter and setting the mother's weight to zero), and the overall set of start vectors is still representative of the steady state dynamics of the system.

Instead of the straight replacement of one mother start vector by one daughter start vector, the simulation is performed multiple times for each mother start vector; daughter vectors are found for the two neighboring points, together with the branching ratio. The mother vector is replaced by two daughter vectors, randomly chosen from those calculated, with one at each neighboring point; the weight of the mother vector is allocated to the daughters proportionately to the branching ratio.

It is now possible to focus the simulation on higher energies in the confining potential where the atom is found more rarely. Once $M$ mother vectors have been propagated for a point $E_{i}$, it is decided that enough data has been gathered to characterize the behavior of the system around this point, which is called `full'. The start vectors at point $E_{i}$ are excluded from being propagated further using the Monte Carlo method; the choice of mother vector to propagation is now determined from the weights of the start vectors at all points which are not full.

Although no further daughter vectors are generated from full points, weight is still added to full points from neighboring partially-filled points; if this goes uncompensated, it will lead to distortion of the distribution of weights between the points. Therefore as each mother vector is propagated and is replaced by daughter vectors, weight is reallocated amongst the full points, and from full points to partially-filled points, in order to preserve the relative allocation of weights between points. Details of this reallocation are given in Appendix \ref{HMCME}. 

This method of extending the Monte Carlo simulation is not unique; in practice the method outlined was found to give the best balance between reproducibility, accuracy and computation time for the problem in hand.

\subsection{The master equation}
\label{MarkNonMark}

The method outlined in this section (and summarized in Appendix \ref{HMCME}) ensures that, as the simulation is run, it progresses from concentrating on the low energy region to the high energy region where the system is found only rarely. Along the way it has amassed data concerning the branching ratios between neighboring energy points (the probability that a system, starting at one point, will end up at either neighbor) together with the time taken to propagate from one point to the next. It is now simple to calculate transfer rates between the points. These transition rates are used in a master equation to calculate the relative populations of the atoms at various energies:
\begin{equation} \label{MasterEq}
\frac{dN_{n}}{dt}=\sum_{m} (r_{mn}N_{m}-r_{nm}N_{n})
\end{equation}
where $r_{mn}\,$$=\,$$p_{mn}/\tau_{m}$ is the rate of transfer from $m$ to $n$, $\tau_{m}$ the average time spent at $m$, and $p_{mn}$ the branching ratio. 

It remains to find the actual quantity of interest---the hopping rate between sites. The hopping rate of an atom at a particular point is directly extracted from the Monte Carlo data taken at that point; the total hopping rate is calculated as the average hopping rate for all points weighted by the relative populations at those points. The hop process is treated as a `sink' for the atomic population, i.e.\ once the atom has departed the well it is removed from the simulation.

\subsection{Implementation of the algorithm to the polarization gradient cooling of confined atoms}
\label{implem}

A few subtleties were encountered when applying the algorithm to the polarization gradient cooling of confined atoms. Firstly, the potential energy of the system is ill-defined as the system has both quantum and classical properties. The usual energy measure in such a scenario is the expectation value of the energy $E_{k}+\langle{\bm V({\bm r})}\rangle$, where ${\bm r}$ is the semi-classical position of the atom. However, this energy measure does not satisfy the two conditions outlined in Section \ref{measEv}, due to quantum jumps caused by the spontaneous emission of radiation, and by poor correlation between the value of the measure and the probability of site hopping (take the example of an excited state atom stationary at a lattice minimum).

Instead, another energy-type measure was used which has the required properties. Using the kinetic energy as before, the potential energy was taken to be the lowest eigenvalue of the atomic system at the position of the atom. The time evolution of this quantity is continuous; furthermore, it correctly accounts for the combination of velocity and position needed to differentiate the main population of atoms from those which hop between sites. The choice is further justified by noting that atoms undergoing polarization gradient cooling in a sublevel-independent lattice tend to be optically pumped to this lowest energy ground-state sublevel (see Sect.\ \ref{orthlin}).

The energy spacing of the points was in practice determined by a compromise between precision and computation time; about 25 points were used spread over the energy range 0 to 1.3$V_{0}$, where $V_{0}$ is the lattice depth. If more closely-spaced points were used, the computation proceeded faster, but at the expense of a larger spread in the prediction of the hopping rate between wells. This was attributed to insufficient diffusion of the state vector as it is propagated from one point to the next leading to observable `clumping' of the simulated trajectories, and consequently undersampling of the phase space of start vectors at each point. Increasing the spacing between points, and so increasing the time the system spent at any particular point, decreased this trajectory clumping by increasing the influence of diffusive processes over shorter-time non-diffusive processes; consequently this decreased the variation in the predicted jump rate between runs of the HMCME simulation at the expense of increased computation time. Around 150 start vectors were propagated for most points, rising to ten times this number near the lip of the lattice potential. 

The accuracy of the algorithm presented in this section was tested by comparison with Monte Carlo simulations performed conventionally using the method described in Section \ref{MCsimul} (which had been tested against previously published data---see Section \ref{compMeth}). There was excellent agreement between the HMCME method and the conventional Monte Carlo method throughout the region of overlap for all the data presented in this article (e.g.\ Figure \ref{ExPop}). 

\section{The frictional force and energy loss rate}\label{FricSect}

To investigate the effect of the pinning lattice on the polarization gradient cooling of ultracold atoms it is instructive to compare the form of the frictional force with and without the lattice present. Conventionally the frictional force is calculated for an atom on a constant-velocity trajectory \cite{dalibard_laser_1989}; however this does not appropriately describe the motion of atoms moving in an additional lattice potential. The average force and the average energy loss rate are instead calculated for an atom on a constant-energy trajectory.

As has been discussed in Section \ref{implem}, there is ambiguity in the definition of an energy measure in the semi-classical situation. The measure used here for the potential energy (as in the rest of this article) is the position-dependent lowest-energy eigenvalue of the atom subject to both the pinning lattice and cooling light fields. The average force and energy loss rates are calculated by integrating Equations \ref{internal_app} to \ref{motion2} using a density matrix method.

\begin{figure*}[t]
\centering
\hspace{-0.9cm}\subfloat[Without the pinning lattice]{\label{coolFV}\includegraphics[scale=0.37]{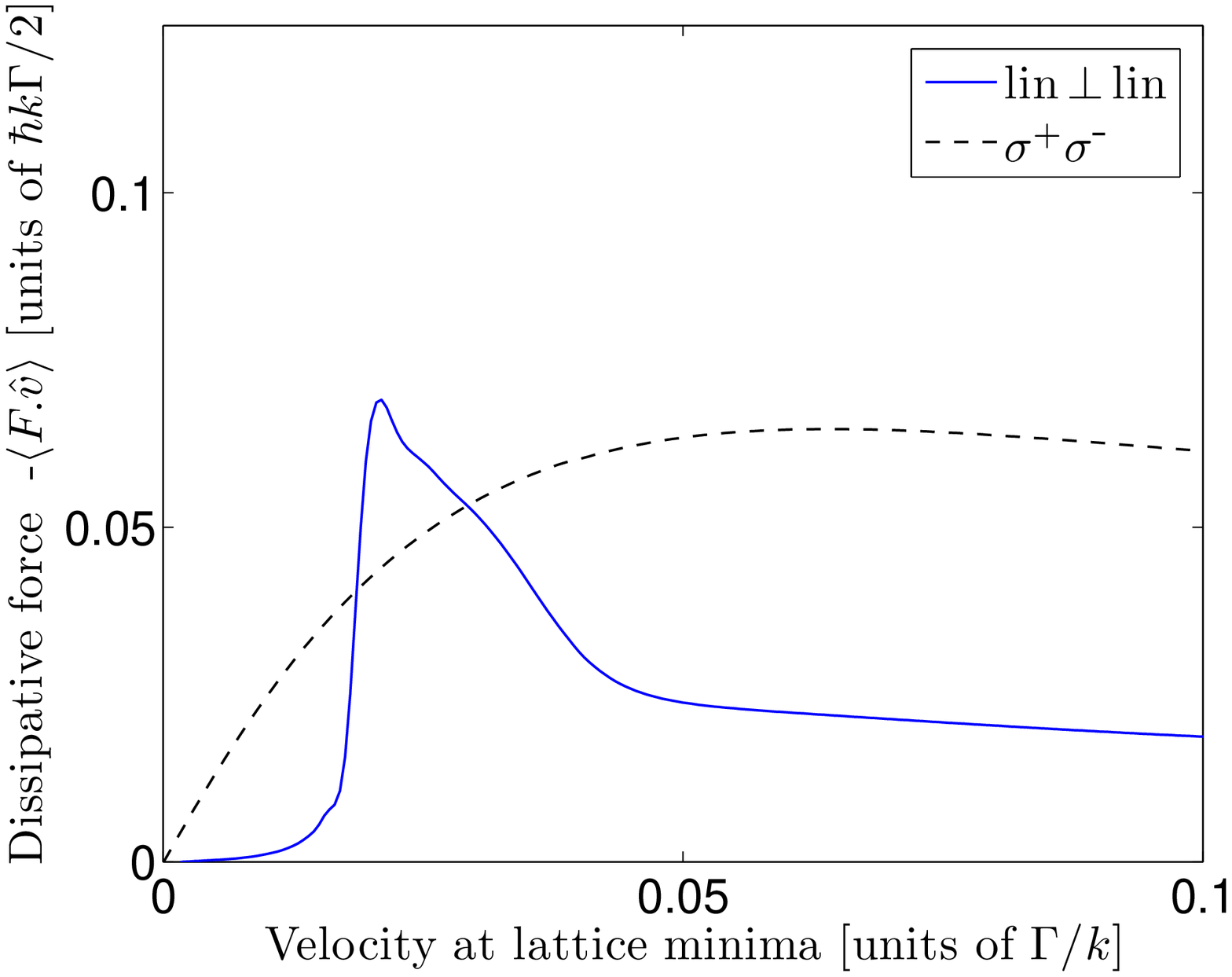}}\hspace{0.5cm}
\subfloat[Without the pinning lattice]{\label{coolFE}\includegraphics[scale=0.37]{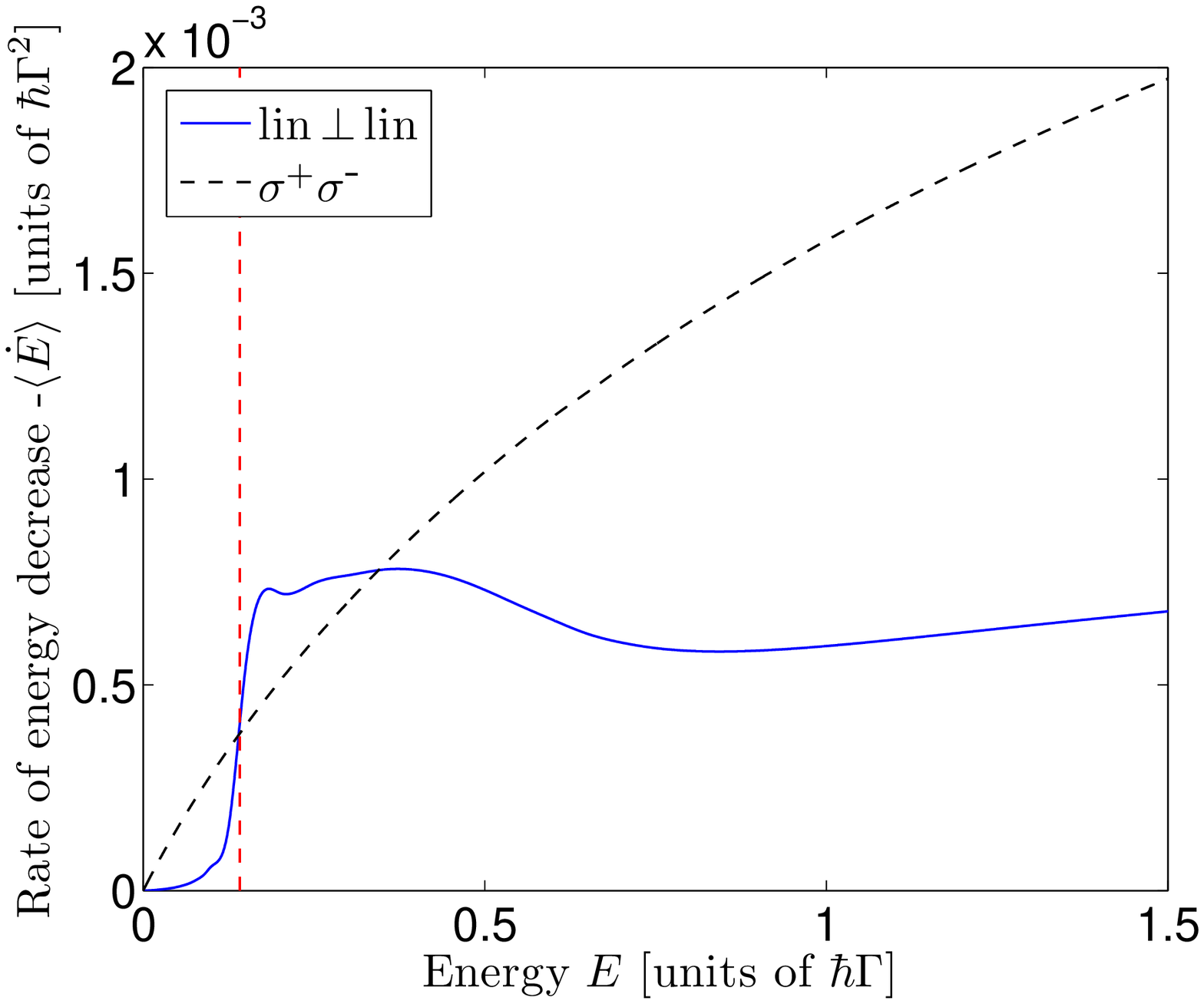}}\\
\centering
\hspace{-0.9cm}\subfloat[In the pinning lattice]{\label{coolTV}\includegraphics[scale=0.37]{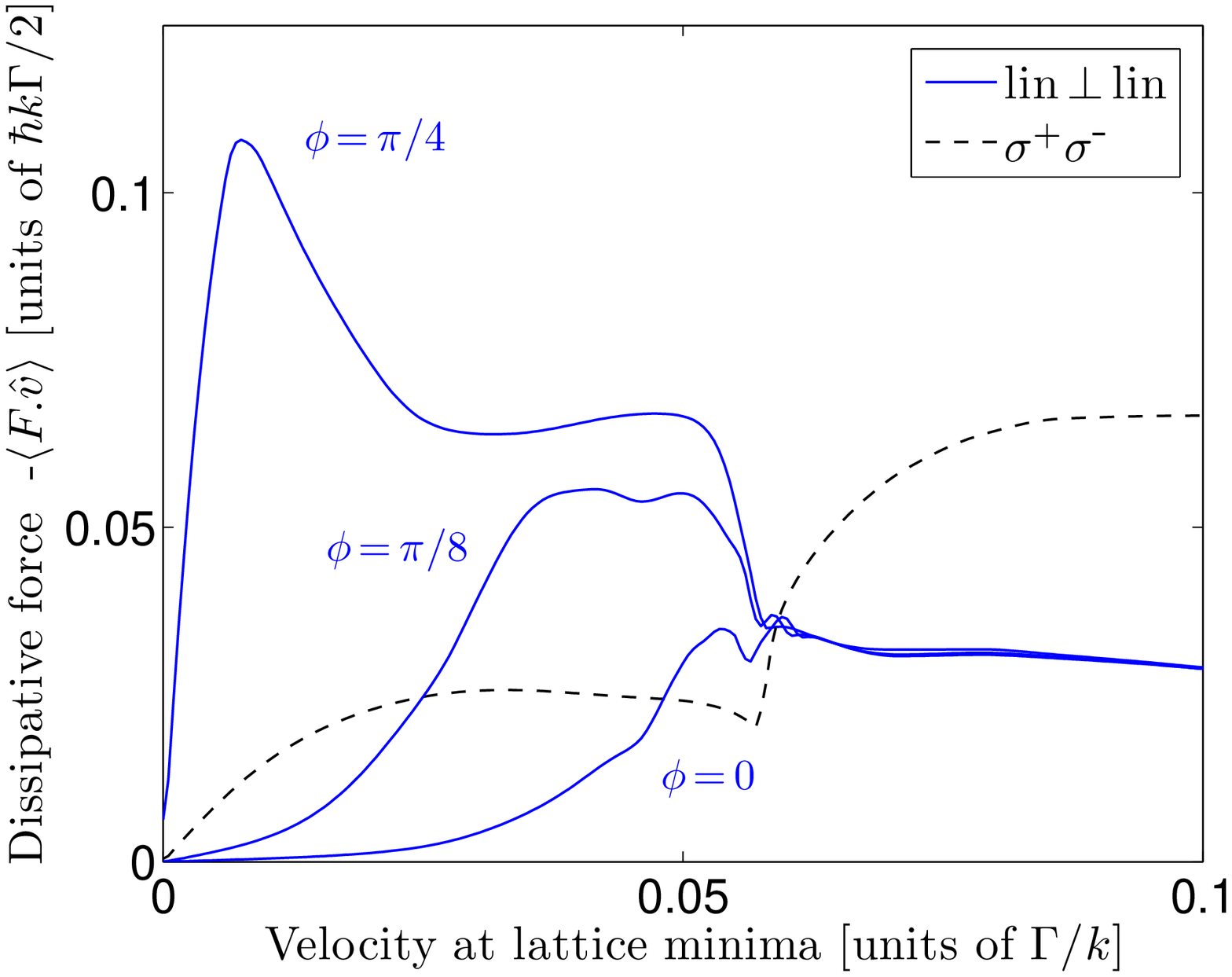}}\hspace{0.5cm}
\subfloat[In the pinning lattice]{\label{coolTE}\includegraphics[scale=0.37]{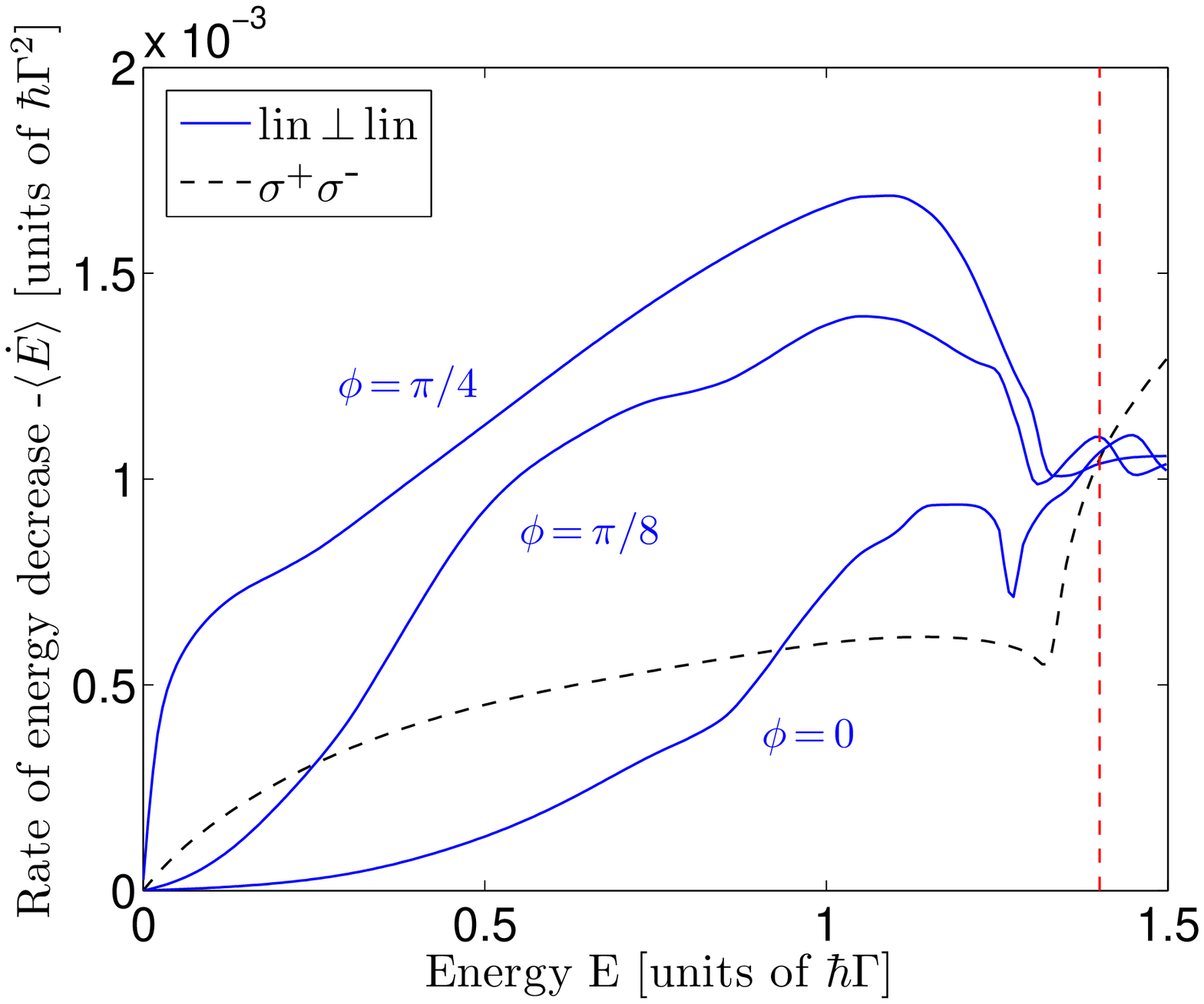}}
\caption{(Color online) Polarization gradient cooling of $^{87}$Rb in one dimension calculated on trajectories of constant energy. (a) Force versus velocity without the pinning lattice. (b) Energy loss rate versus energy without the pinning lattice. (c) Force versus velocity in the pinning lattice. (d) Energy loss rate versus energy in the pinning lattice. The vertical dashed red line indicates the depth of the lin$\,{\perp}\,$lin cooling light lattice in Figure (b) and the depth of the bare pinning lattice (i.e. in the absence of cooling light) in Figure (d). The intensity of each molasses beam is $10.8\,$mWcm$^{-2}$ with frequency detuning -4$\Gamma$ from the D$_{2}$ line. The pinning lattice potential with depth 1.4$\hbar\Gamma$ is generated by counter-propagating linearly polarized beams with frequency detuning $+$2000$\Gamma$ from the D$_{1}$ line. The parameter $\Gamma$ is the angular frequency width of the D$_{2}$ line.}
\end{figure*}

\subsection{No additional lattice present}

The force and energy loss rates for polarization gradient cooling without an additional lattice present are presented in Figures \ref{coolFV} and \ref{coolFE}. The dependence of the force on velocity for the $\sigma^{+}\sigma^{-}$ configuration is very similar to previous analyses \cite{dalibard_laser_1989}, while the lin$\,{\perp}\,$lin configuration looks somewhat different. 

The differences for the lin$\,{\perp}\,$lin case are due to performing the calculation on a constant-energy rather than constant-velocity trajectory. For energies below 0.14$\,\hbar\Gamma$ the atom is confined to a single site of the optical lattice induced by the cooling light (see Fig.\ \ref{nextCool1}), so the atom is not cooled efficiently. The atom experiences stronger cooling once it can move along the lattice. On the other hand, the induced dipole potential is spatially homogeneous in the $\sigma^{+}\sigma^{-}$ configuration, so atoms moving on trajectories which have constant energy also have constant velocity; results are found which are very similar to previous analyses.

If an energy measure is used which does not contain potential energy contributions from the cooling light, the form of the lin$\,{\perp}\,$lin velocity dependence alters to closely resemble that found in previous analyses (e.g.\ \cite{dalibard_laser_1989}). It is worth noting that in practice atoms will follow neither constant-velocity nor constant-energy trajectories.

\subsection{With additional lattice present}
\label{lat_pres}

The cooling force and energy loss rate for atoms in a pinning lattice of depth 1.4$\hbar\Gamma$ are shown in Figures \ref{coolTV} and \ref{coolTE}. The velocity plotted in Figure \ref{coolTV} is the velocity of an atom at the minima of the lattice potential. 

The frictional energy loss profile for bound atoms is dependent on phase of the cooling beams at the pinning lattice site in the lin$\,{\perp}\,$lin configuration but not in the $\sigma^{+}\sigma^{-}$ configuration. As discussed in Section \ref{orthlin}, the cooling is very inefficient for atoms in a lattice with a $\phi\,$=$\,0$ phase relative to the cooling light, but is efficient in the $\phi\,$=$\,\pi/4$ case.

The character of the cooling is different for atoms which have an energy greater than the depth of the lattice. The cooling of unbound atoms is similar to the cooling of atoms without the lattice present---the lin$\,{\perp}\,$lin configuration becomes phase independent, while the $\sigma^{+}\sigma^{-}$ configuration has greater cooling power. On the other hand, atoms with energies around the lattice binding energy experience reduced cooling power. This is due to time-averaging; these atoms travel slowly when near the lip of the potential, so they are cooled more weakly at that time, and the average force falls. These dips occur at slightly different positions due to the different depths of the overall potential in each case. 

\section{Simulated in-lattice polarization gradient cooling}
\label{ExampDyn}
\subsection{Simulated scenarios}
\label{SimScen}
As discussed in Section \ref{cool3dinit}, three-dimensional molasses necessarily contains sub-wavelength scale intensity and polarization gradients. If an atom is confined so that it only samples the light field in a sub-wavelength region, the cooling efficiency is heavily dependent on the relative phases of the molasses light (as demonstrated in Section \ref{lat_pres}). Experiments requiring reliably efficient three-dimensional cooling for tightly confined atoms need to find a way to overcome this problem.

Two experimental scenarios are simulated in this section; they differ in the method by which the phase problem is tackled. One method (which will be called the 1D alternating configuration) uses a one-dimensional $\sigma^{+}\sigma^{-}$ cooling beam configuration which is alternated between the three axes in turn. This can be realized in an experiment by using square-wave pulses to modulate the input to three acousto-optical modulators. In the other method (the 3D offset configuration), all six $\sigma^{+}\sigma^{-}$ cooling beams are used at once, with a small frequency difference between each set of counter-propagating beams. This sweeps the relative phases of the cooling light at each site. The initial values of the relative phases of the cooling light at the pinning lattice site were chosen to be random for each run of the HMCME simulation.

The commonly used species $^{87}$Rb is chosen for the analysis; similar results are expected to hold for other atomic species which are subject to polarization gradient cooling. The pinning lattice was chosen to be near the D$_{1}$ atomic transition to allow efficient filtration of the intense lattice light from the weaker fluorescence light during the measurement. The pinning lattice frequency detuning is chosen to be much greater than the hyperfine structure of the D$_{1}$ line, and the lattice polarizations chosen to be linear, in order minimize the potential energy differences between the ground-state sublevels (see Section \ref{orthlin} and \cite{winoto_laser_1999,nelson_imaging_2007}).

\begin{table*}[t]
\centering
\caption{Parameters for the scenarios discussed in Section \ref{ExampDyn}. The frequency detunings are given with respect to the closed $F\,$=$\,$2$\,\rightarrow\,$$F'\,$=$\,$3 transition for the D$_{2}$ cooling light and from the $F\,$=$\,$2$\,\rightarrow\,$$F'\,$=$\,$2 transition for the D$_{1}$ pinning lattice light. The flash cycle time is defined as three times the cooling flash duration in any one direction. The frequency offset is the defined as the difference in frequency between the three sets of cooling beams ($\delta\,$=$\,\nu_{x}\,$-$\,\nu_{y}\,$=$\,\nu_{y}\,$-$\,\nu_{z}$=$\,(\nu_{x}\,$-$\,\nu_{z})/2$). The parameter $\Gamma$ is the angular frequency width of the D$_{2}$ line. The pinning lattice depth is given in Kelvin as $E/k_{B}$.}
\begin{tabular}{llcc}
\hline
\hline
Atomic species&Species& \multicolumn{2}{c}{$^{87}$Rb} \\
\hline
\multirow{8} {*}{Pinning lattice} & Line & \multicolumn{2}{c}{D$_{1}$ (794.98$\,$nm)} \\
& Frequency detuning & \multicolumn{2}{c}{$+$2000$\Gamma$} \\
& Intensity per beam & \multicolumn{2}{c}{3.0$\,\times\,$$10^{4}\,$mWcm$^{-2}$} \\
& Depth & \multicolumn{2}{c}{1.4$\hbar\Gamma$ (408$\,\mu$K)} \\
& Period & \multicolumn{2}{c}{397$\,$nm} \\
& \multirow{2} {*}{Character} & \multicolumn{2}{c}{1D counter-propagating in each} \\
& & \multicolumn{2}{c}{$\;$ direction (intensities add)}\\
& Polarization & \multicolumn{2}{c}{Linear} \\
\hline
\multirow{8} {*}{Cooling light} & Line & \multicolumn{2}{c}{D$_{2}$ (780.24$\,$nm) $F\,$$=\,$$2\,$$\rightarrow \,$$F'\,$$=\,$$3$}\\
& Frequency detuning & \multicolumn{2}{c}{-4$\Gamma$}\\
& Intensity per beam & $10.8\,$mWcm$^{-2}$ & $1.81\,$mWcm$^{-2}$\\
& Character & Alternating 1D $\sigma^{+}\sigma^{-}$ & Offset 3D $\sigma^{+}\sigma^{-}$\\
& Flash duration & 18$\,$\textmu s & \textrm{---} \\
& Flash cycle frequency & 18$\,$kHz & \textrm{---} \\
& Frequency offset & \textrm{---} & 6.1$\,$kHz \\
\hline \hline
\end{tabular}\label{ParamsEx}
\end{table*}

\begin{table*}[t]
\centering
\caption{Data from simulations using the parameters given in Table \ref{ParamsEx}. The quoted site lifetimes are due to radiative heating only, and do not include the effect of background gas collisions. The quoted errors on the calculated parameters are the one standard deviation random errors and do not include systematic effects. The temperature calculations include kinetic and potential energy contributions.}
\begin{tabular}{ccc}
\hline \hline Cooling configuration & Alternating 1D $\sigma^{+}\sigma^{-}$ & Offset 3D $\sigma^{+}\sigma^{-}$\\\hline 
log$_{10}$(Site lifetime/seconds)& 6.5$\pm$0.3 & 6.6$\pm$1.6 \\
Mean site lifetime& 3.3$\,\times\,$$10^{6}\,$s & 3.8$\,\times\,$$10^{6}\,$s\\
Mean temperature& 8.5$\pm$0.3$\,$\textmu K & 10.3$\pm$2.0$\,$\textmu K \\
Mean scattering rate& (1.89$\pm$0.001)$\,\times\,$$10^{6}\,$s$^{-1}$& (1.41$\pm$0.11)$\,\times\,$$10^{6}\,$s$^{-1}$ \\
\hline
\hline
\end{tabular}\label{ParamsOut}
\end{table*}

\subsection{Discussion of the results}
\label{discussSimDyn}

\begin{figure*}[t]
\centering
\subfloat[]{\vspace{3cm}\label{ExRates1}\includegraphics[scale=0.405]{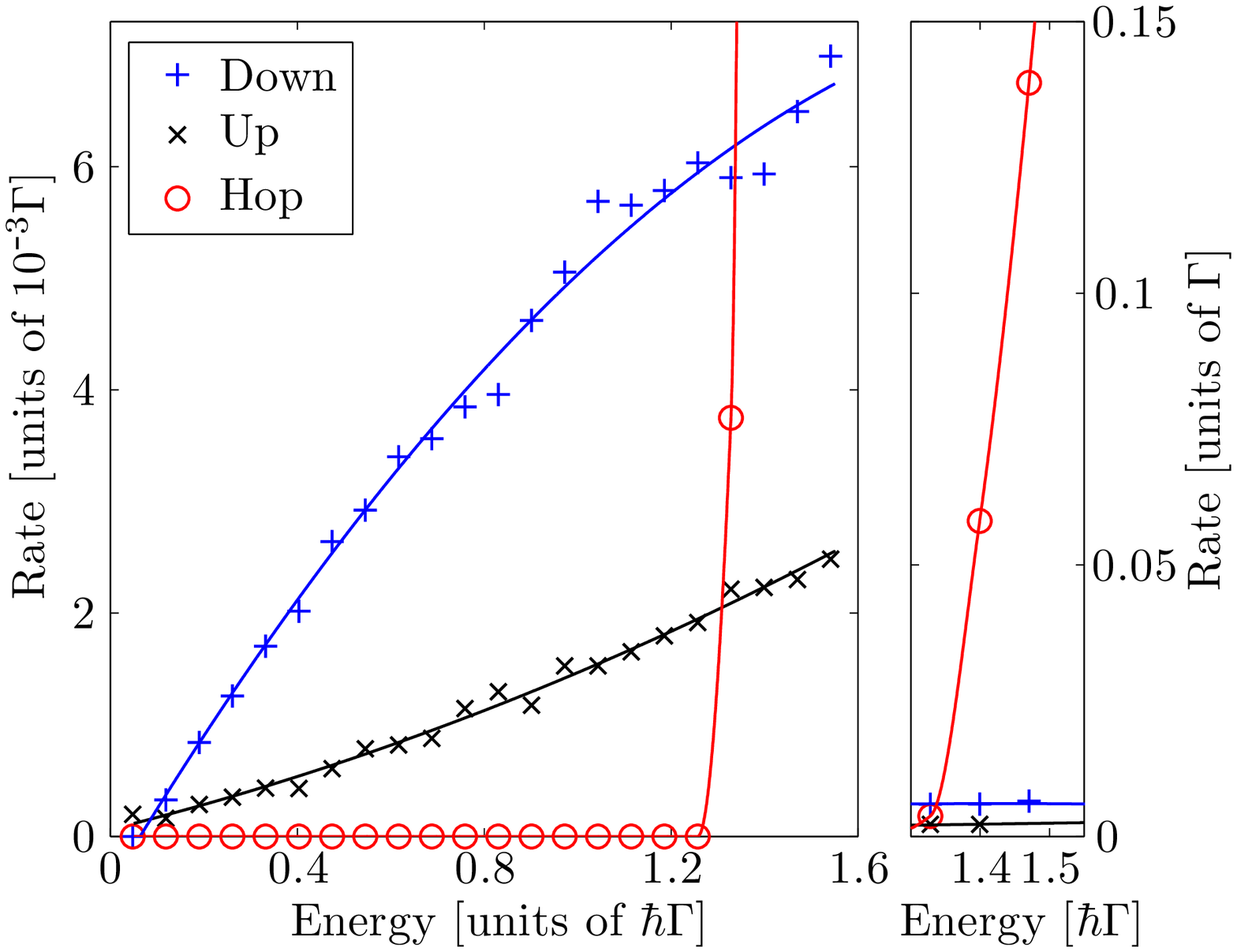}}\hspace{0.1cm}
\subfloat[]{\label{ExPop}\includegraphics[scale=0.40538]{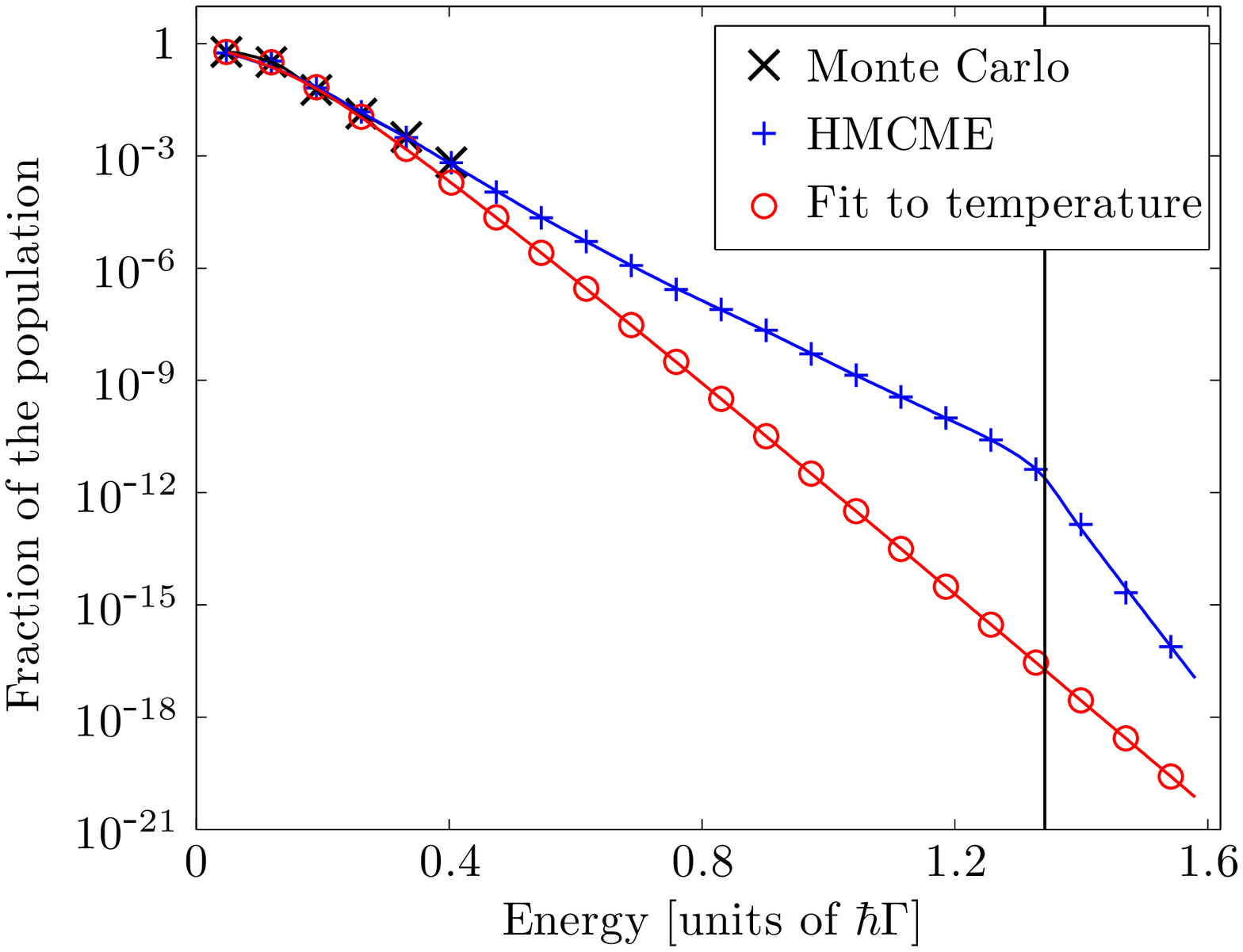}}
\caption{(Color online) (a) Transition rates between different energy classes, as found from a single run of the HMCME simulation. Atoms are categorized according to energy according to the method described in Section \ref{ext_MCSect} and Appendix \ref{HMCME}. Rates are given for transfer into the next higher or lower energy category together with the hop rate for transfer to a neighboring well. The data was generated using the HMCME method using the parameters of Table \ref{ParamsEx} in the alternating 1D cooling configuration. (b) Histogram of population versus energy. The populations for the HMCME technique were calculated from the rates given in Figure (a); the Monte Carlo populations were found directly from a conventional Monte Carlo simulation. The depth of the potential, marked with the vertical black line, is 1.34$\hbar\Gamma$, which is less than the bare pinning lattice (1.4$\hbar\Gamma$) due to the dressing with the cooling light.}\label{ExPlots}
\end{figure*}

The results of the HMCME simulation are shown in Figure \ref{ExPlots} using the parameters given in Table \ref{ParamsEx} for the alternating 1D cooling configuration. Figure \ref{ExRates1} gives the transition rates between points, while Figure \ref{ExPop} gives the steady state populations at each energy point i.e.\ the proportion of the total population which last encountered that specific energy point rather than any other.

The transfer rate to lower energy is greater than the transfer rate to higher energy for all but the lowest-lying points, i.e.\ on average the atom is cooled at all but the lowest energies. The difference in the rates increases with energy; this is to be expected as the average frictional energy loss increases with energy throughout this region (see Figure \ref{coolTE}). It is seen that the rate datum points have a small spread around the trend line due to the statistical uncertainty implicit in the Monte Carlo method. This leads to some variation in the site lifetime between simulation runs, and is reflected in the quoted statistical uncertainty on the lifetime (Table \ref{ParamsOut}).

The `hop' rate for a certain energy class is the rate at which atoms transfer from that energy class to a neighboring site; this dominates the behavior of atoms which have energy above the binding energy of the lattice. Only atoms with energy slightly above the lattice binding energy have an appreciable probability in this configuration of being captured by one particular site; atoms with larger energies will move between many sites before capture, or may leave the lattice. The effects of the jump process are seen in Figure \ref{ExPop} in the sharp decrease of the population with energies above the lip of the lattice (it is assumed that, once the atom has hopped, it does not return to the original well). Note that the quantum mechanical tunneling of (deeply bound) atoms between lattice sites is negligible due to the substantial depth of the lattice.

There is an excellent fit between the population predicted by the standard Monte Carlo simulation and that predicted by the HMCME simulation in the region in which the standard Monte Carlo simulation can be used. That the HMCME method simulates the system well in a known region gives good confidence that the HMCME method provides an accurate extrapolation of the dynamics to higher energies.

Assuming ergodicity, the population data is fitted to a temperature once the density of states at the lattice site has been calculated using
\begin{equation}
D(E)=\frac{(2m)^{\frac{3}{2}}}{(2\pi)^{2}\hbar^{3}}\int_{V({\bm r})\leq E}\sqrt{E-V({\bm r})}d^{3}{\bm r}\;.
\end{equation}
The fitted temperature population distribution (essentially the same when fitted to either the HMCME or Monte Carlo data) is plotted in Figure \ref{ExPop}. The fit is accurate in the low energy region, but the full HMCME method predicts more atoms in the tail of the distribution than would be expected at a given temperature. This is due to the description of the population in terms of a temperature being valid only when the frictional force divided by the diffusion coefficient is a linear function of velocity (see for example \cite{metcalf_laser_1999} p66). The frictional force is in fact not linear---the slope of the $\sigma^{+}\sigma^{-}$ force decreases with increasing energy (Fig.\ \ref{coolTV}), so the population in the tail of the distribution is higher than what the temperature of the low-lying atoms would predict. 

The corresponding plots for the 3D offset configuration are similar to those displayed for the 1D alternated cooling configuration. Calculated quantities---the temperature, fluorescent scattering rate and site lifetime due to radiative processes---for both configurations are given in Table \ref{ParamsOut}. The mean time before site hop is defined as the average time for an atom at the steady-state temperature at a pinning lattice site to leave that lattice site, with random phases of the cooling beams at the lattice site. Over a million photons are scattered per second in either scenario, with very long site lifetimes predicted. Of course, such long lifetimes will not be observed in experiments due to background gas collisions; the significance of these figures will be discussed in Section \ref{discuss}. 

The quoted uncertainty in Table \ref{ParamsOut} is the standard deviation on the mean of these quantities over a sample of 100 runs of the complete HMCME simulation; it is a combination of the uncertainty due to the random nature of the Monte Carlo simulation together with a contribution which depends on the relative phases of the cooling beams at the pinning lattice site, as will be discussed in Section \ref{1D3D}. The averages and standard deviations of the hopping rate are calculated in the logarithm throughout this article; this has a similar functional dependence to an average over the temperature. While this uncertainty on the mean hopping rate is large in percentage terms, it is negligible compared to the rate of background gas collisions. The greatest systematic uncertainty in the calculation is likely to arise from the use of the semi-classical approximation, which although necessary to make the problem tractable, is only approximately fulfilled; the parameter $\epsilon_{1}=\hbar k_{R}/\Delta p$, assumed to be much less than unity in this analysis (Sect.\ \ref{semiclass_approx}), is 0.19 for the current example, with the atom scattering around 5 photons per oscillation period. It is worth noting that the high ratio of the dephasing event rate (i.e.\ the photon scatter rate) to the oscillation frequency makes a simulation based on transitions between pure quantum vibrational states problematic. To improve on the semi-classical approximation a fully spatially-dependent density matrix description is needed, which would be very difficult computationally. 

\section{Lifetime before hop versus experimental parameters}
\label{lifeHop}

\begin{figure*}[ht!]
\centering
\subfloat[1D alternating]{\hspace*{0.2cm}\label{VarInten1D}\includegraphics[scale=0.33]{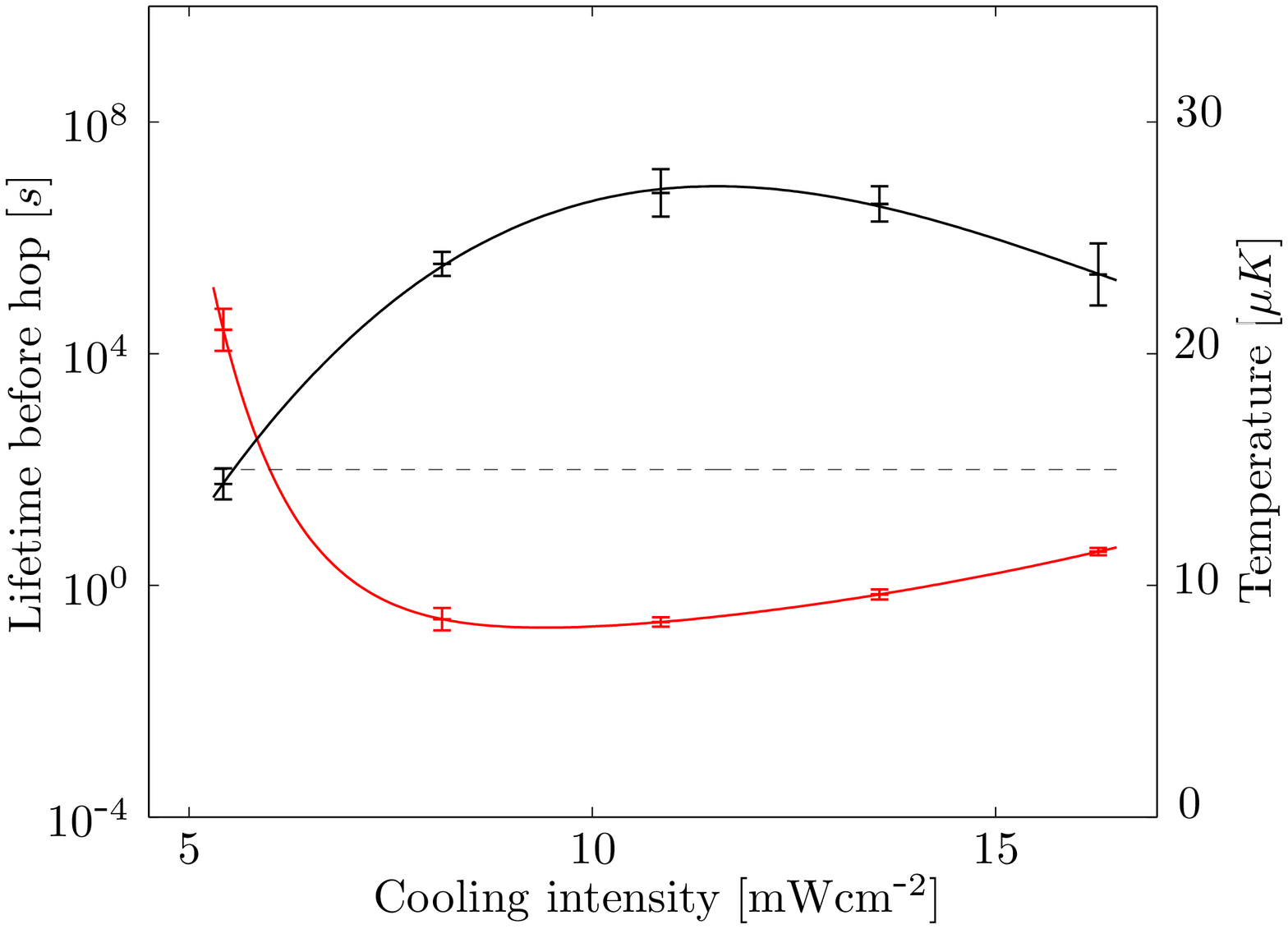}}\hspace*{0.25cm}
\subfloat[3D offset]{\label{VarInten3D}\includegraphics[scale=0.33]{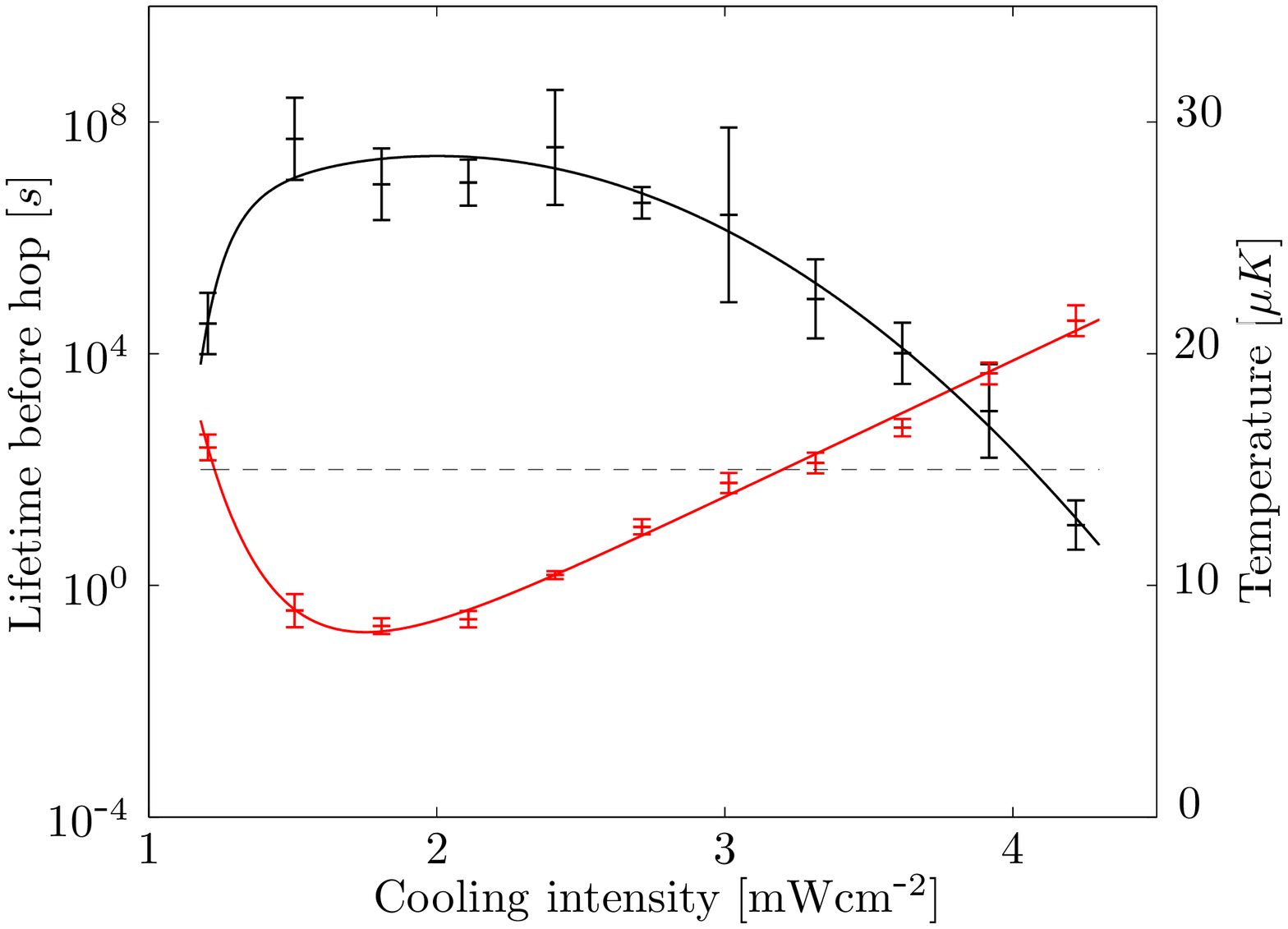}}\\
\subfloat[1D alternating]{\vspace{3cm}\label{VarCoolDet}\includegraphics[scale=0.33]{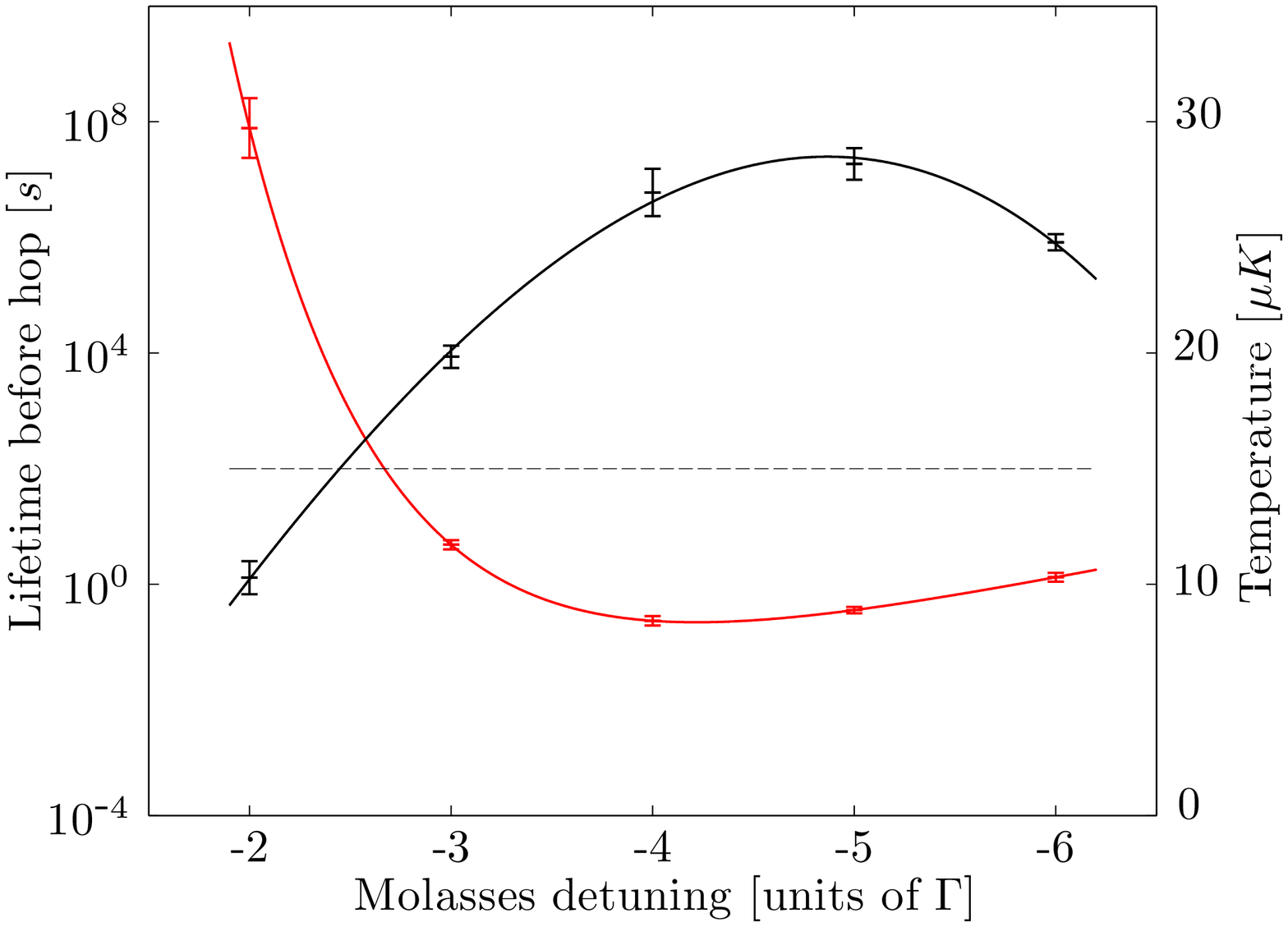}}\hspace*{0.25cm}
\subfloat[3D offset]{\label{VarCoolDet3D}\includegraphics[scale=0.33]{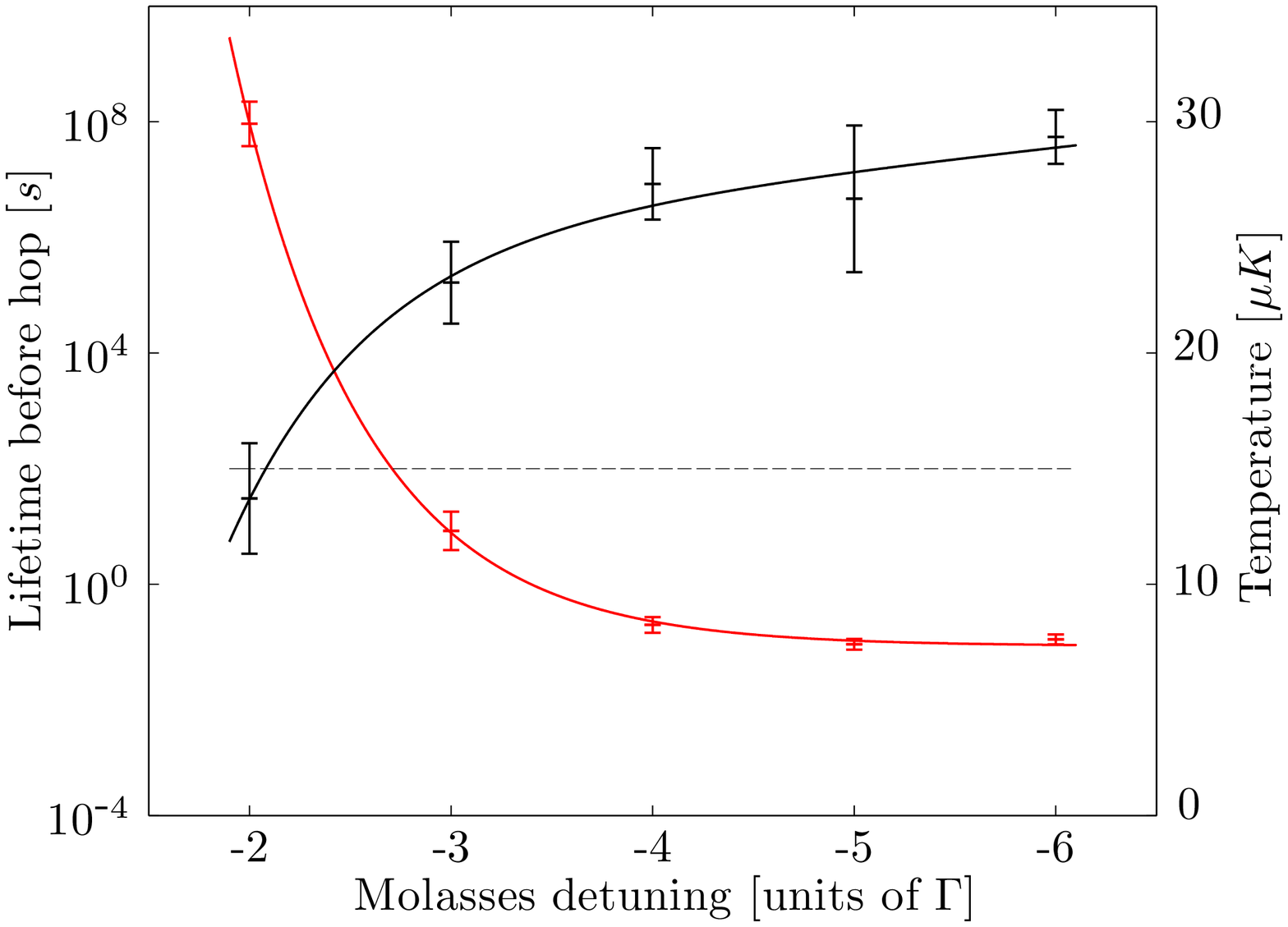}}\\
\subfloat[1D alternating]{\vspace{3cm}\hspace*{0.2cm}\label{VarDepth}\includegraphics[scale=0.33]{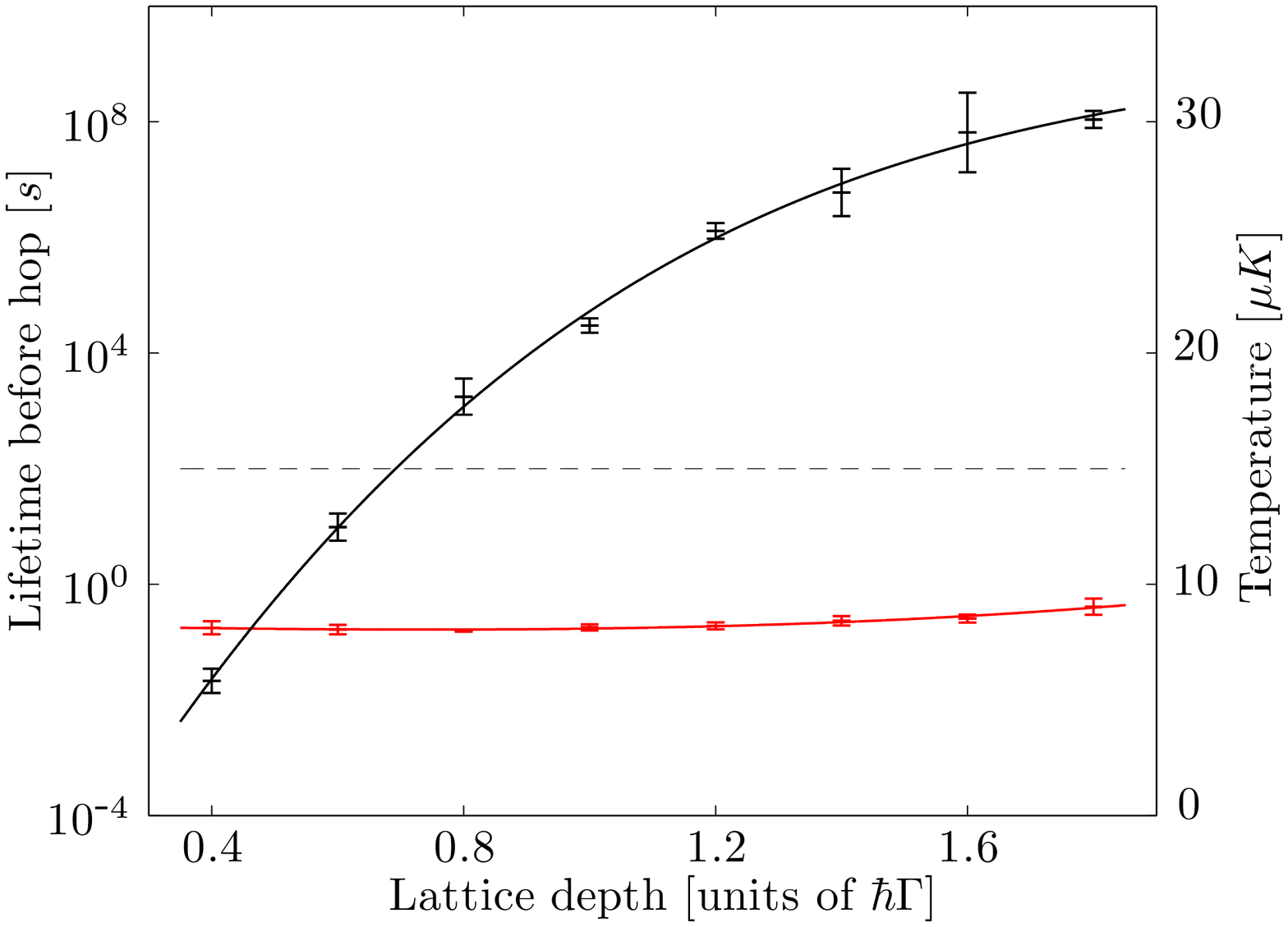}}\hspace*{0.25cm}
\subfloat[3D offset]{\label{VarDepth3D}\includegraphics[scale=0.33]{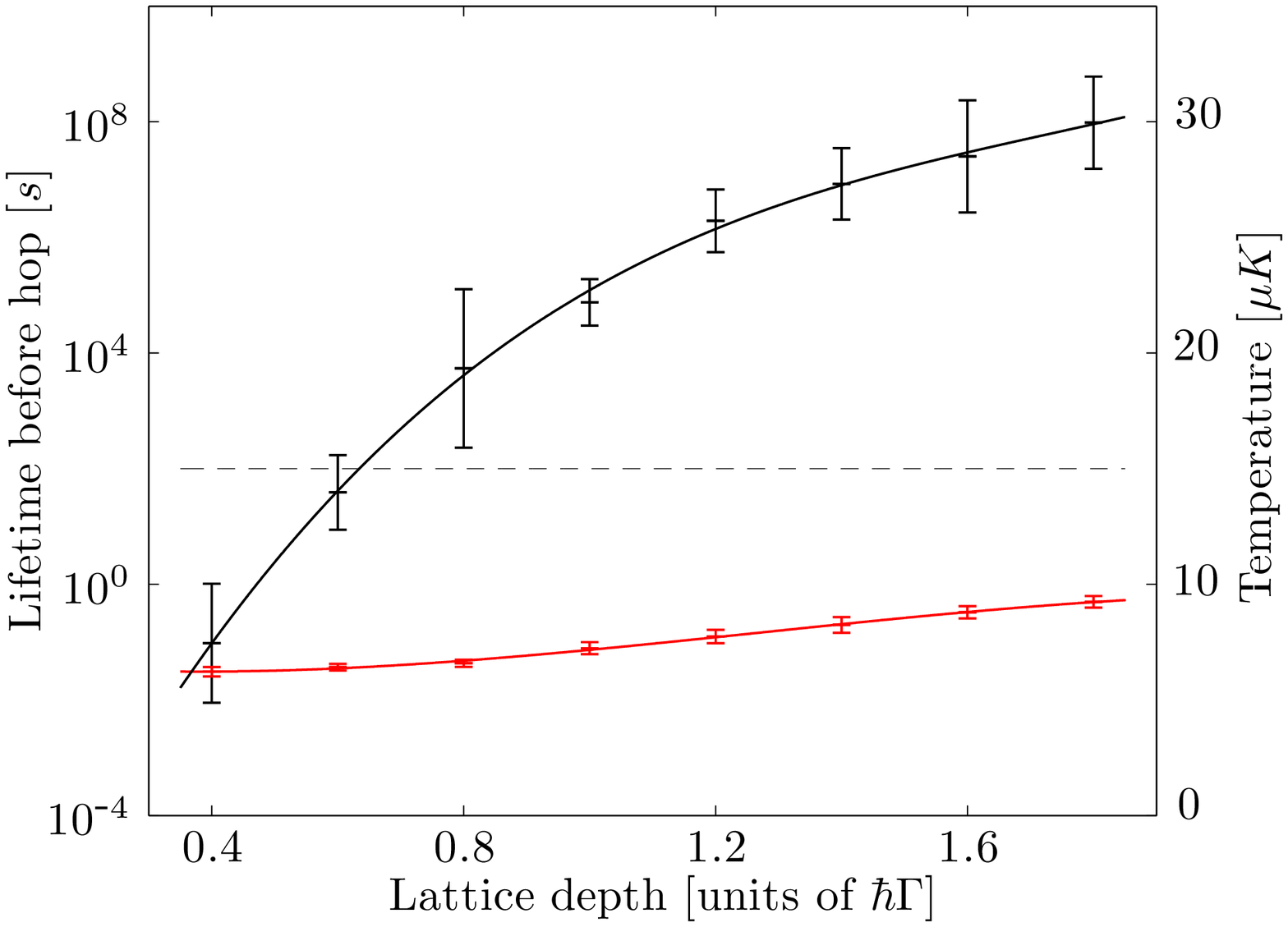}}\\
\subfloat[1D alternating]{\vspace{3cm}\hspace*{0.2cm}\label{VarLattAngle}\includegraphics[scale=0.33]{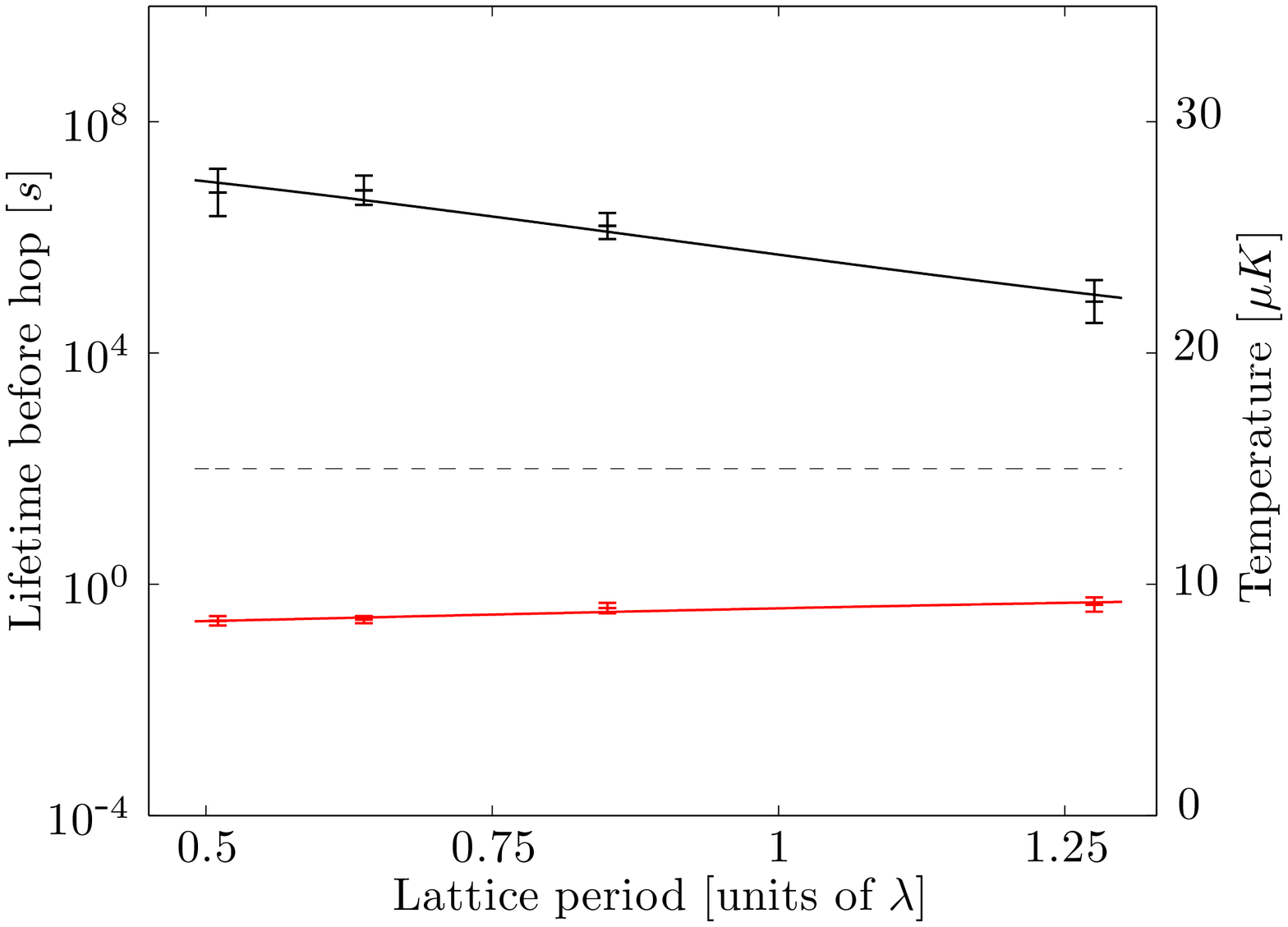}}\hspace*{0.25cm}
\subfloat[3D offset]{\label{VarLattAngle3D}\includegraphics[scale=0.33]{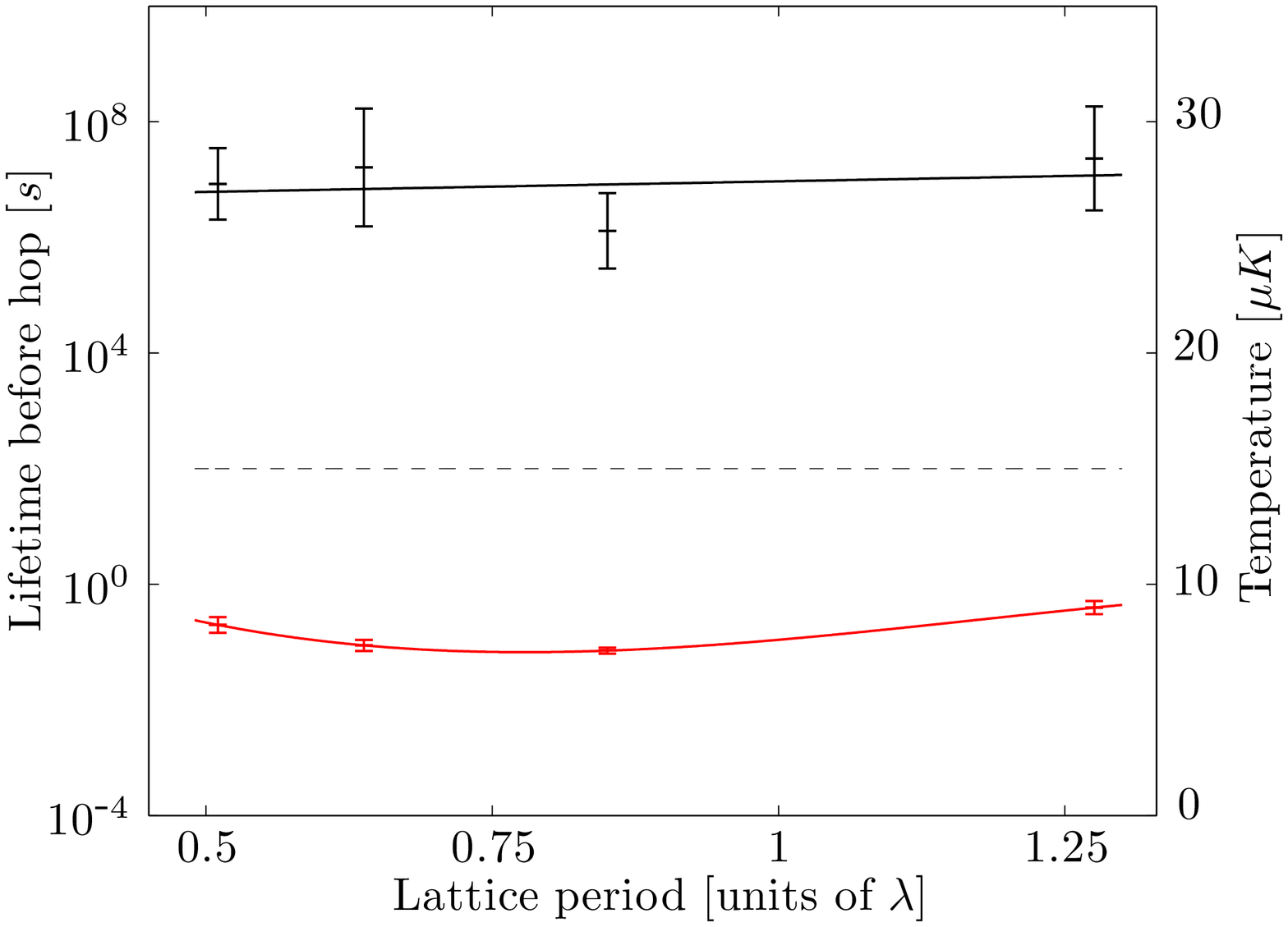}}
\vspace{1cm}
\end{figure*}
\setcounter{figure}{3}
\begin{figure*}[ht!]
\centering
\setcounter{subfigure}{8}
\subfloat[1D alternating]{\vspace{3cm}\label{VarFlash1D}\includegraphics[scale=0.33]{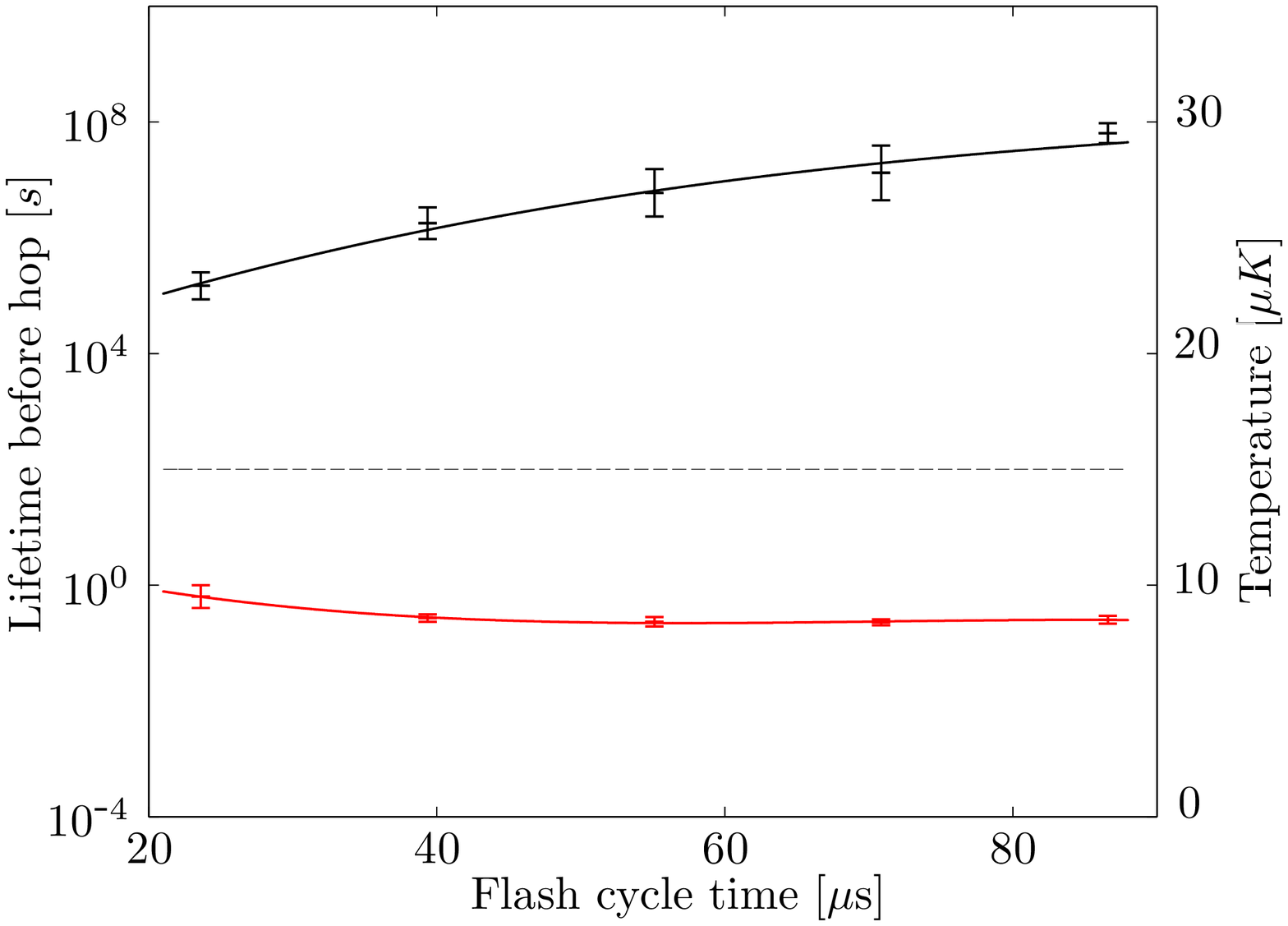}}\hspace*{0.25cm}
\subfloat[3D offset]{\label{VarFlash3D}\includegraphics[scale=0.33]{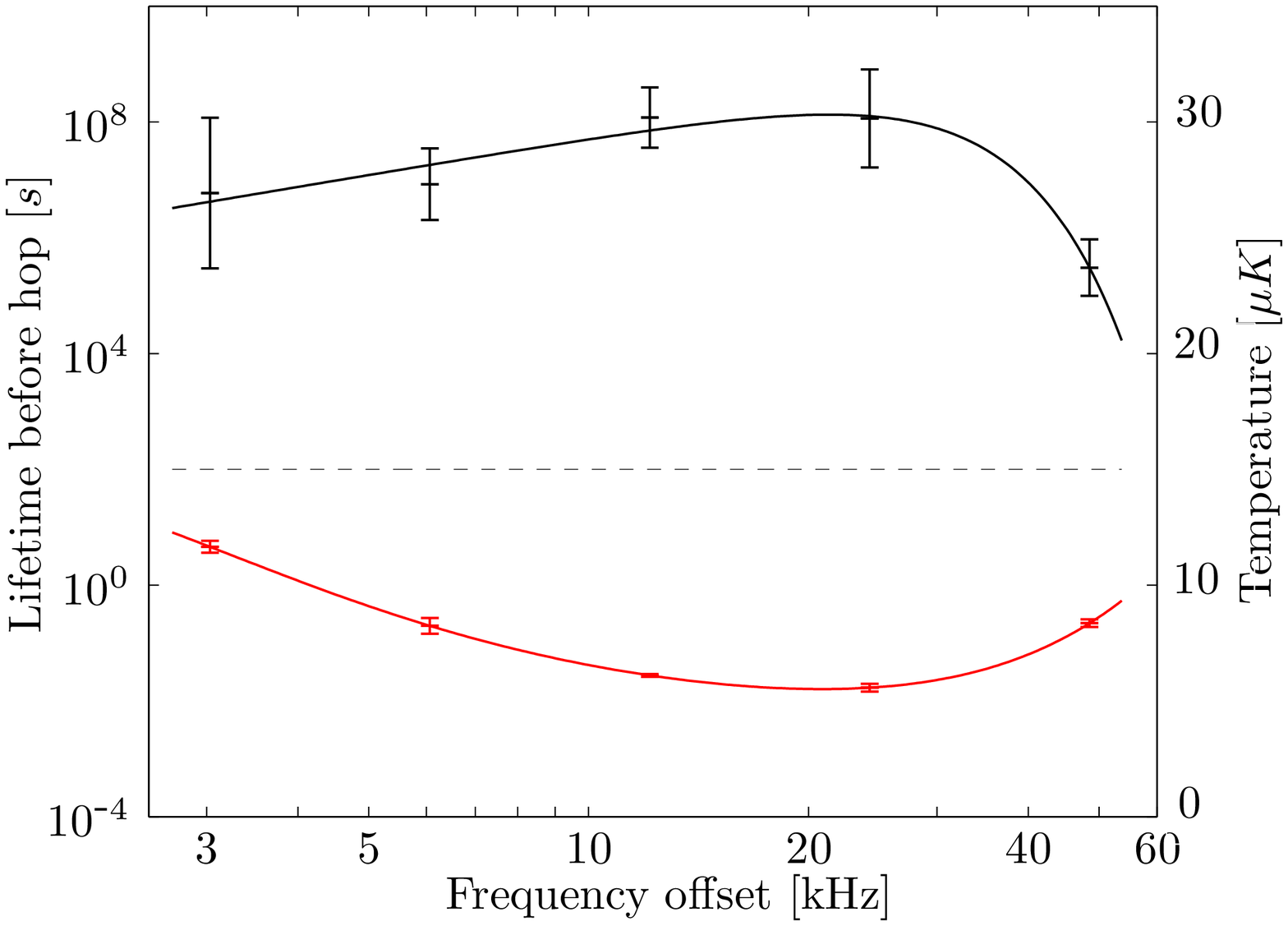}}
\caption{(Color online) The variation of the temperature (lower red line) and site lifetime (upper black line) against experimental parameters for atoms undergoing polarization gradient cooling in a three-dimensional optical lattice. All of the parameters apart from the parameter being varied are given in Table \ref{ParamsEx}, except for the cooling intensity in Figures (c) and (d), which was varied along with the cooling light frequency detuning in order to maintain a constant saturation parameter. The phases of the cooling beams relative to the pinning lattice site were the same for all data. Five runs of the HMCME simulation were used for each data point; the sample standard deviation is displayed. For comparison, the dotted horizontal line indicates the approximate lifetime of the atom at the site due to background gas collisions.}
\label{VarStuff}
\end{figure*}

The dimensionality of the phase space of the experimental parameters is fairly large, so a complete characterization of the system is not feasible within the scope of this article. Instead, the dependence of the hopping rate on individual parameters is investigated. One parameter at a time will be varied, with the other parameters of the system held fixed at the values given in Table \ref{ParamsEx}, which have been chosen to give good performance. Both the 1D alternating and 3D frequency offset molasses configurations, as discussed in Section \ref{SimScen}, will be investigated. The relative phases of the cooling beams at the lattice site are not varied in this section (as this would greatly increase the number of computations needed). The effect of the relative phases will be discussed separately in Section \ref{1D3D}.

In fact, nearly all of the parameter sets presented in this section are predicted, in experiments, to have a site lifetime limited by the rate of background gas collisions rather than radiative processes; this will be discussed in Section \ref{discuss}. The objective of this section, however, is a comparative study of the efficiency of polarization gradient cooling in a deep optical lattice; the quoted site hopping rate, although for most parameters unobservable in the physical system, in this case serves as a useful proxy measure for the cooling efficiency.

The dependence of the temperature and the site hopping rate on the intensity of the molasses beams is shown in Figures \ref{VarInten1D} and \ref{VarInten3D}. For both the 1D alternating and 3D offset configurations the temperature passes through a minimum in the range investigated. At higher intensities, the temperature displays a linear dependence on intensity, the same dependence as for cooling without the lattice present \cite{dalibard_laser_1989,salomon_laser_1990}. The temperature increases markedly for low cooling intensities. It is suggested, as discussed in Section \ref{orthlin} and in Reference \cite{nelson_imaging_2007}, that the behavior at low intensity is caused by less efficient following of the local equilibrium population by the atom in its motion when the optical pumping rate is lower, leading to a corresponding decrease in the efficiency of the cooling. The intensity at which the atoms are at minimum temperature is greater than for the case without the lattice present (for similar frequency detuning---c.f.\ for example \cite{salomon_laser_1990}). The time before the atom is lost from the site displays a complimentary relationship with the temperature, as expected. 

Figures \ref{VarCoolDet} and \ref{VarCoolDet3D} display the results of varying the molasses frequency detuning at constant saturation of the atomic transition. The saturation was kept constant in order to maintain an approximately constant scattering rate (and so an approximately constant optical pumping rate) as the frequency detuning is varied. To achieve this the intensity of the cooling beams were varied along with the frequency detuning $\Delta$ in order to keep the saturation parameter $\Omega^{2}/2(\Delta^{2}\,$+$\,\Gamma^{2}/4)$ constant.

The standard theoretical model of polarization gradient cooling for a $J\,$=$\,1$ transition \cite{dalibard_laser_1989} indicates that the temperature varies proportional to $(\alpha^{2}/\Delta) + \Delta$ with $\alpha$$\,\approx\,$3$\Gamma$, with the transition point $\alpha$ between the two regimes dependent on the Clebsch-Gordon coefficients of the transition. In fact this form of dependence matches the data in the confined case reasonably well, albeit for a somewhat higher transition point. Again, the site hopping lifetime displays a complimentary relationship with the temperature.

As expected, the site hopping rate depends strongly on the pinning lattice depth (Figures \ref{VarDepth} and \ref{VarDepth3D}). If the depth of the lattice did not affect the cooling dynamics, the site hopping rate would be expected to exhibit an exponential dependence on the lattice depth. The line on the logarithmic plot is in fact not straight but has a negative second derivative, indicating that the cooling is less effective at larger pinning lattice depths; this is also seen in the slight increase of temperature with lattice depth. It is suggested that two effects may contribute to this. Firstly the oscillation cycle time decreases with lattice depth; the cooling efficiency is degraded, as discussed above, when there are too few spontaneous scattering events per oscillation cycle. Secondly, the ground sublevel-dependent component of the lattice potential increases with the lattice depth; this differential in general disrupts the cooling mechanism (as discussed in Section \ref{orthlin}), disproportionatly affecting high-energy atoms which travel far from the lattice minimum.  Nevertheless, in the given parameter range these effects are not enough to generate a net increase in the site hopping rate as the lattice depth is increased. 

The results of changing the spatial period of the lattice by altering the angles between the lattice beams are shown in Figures \ref{VarLattAngle} and \ref{VarLattAngle3D}. In the range investigated (lattice spatial periods of 400$\,$nm to 1000$\,$nm), the temperatures remain approximately constant, as does the site hopping rate for the case of the 3D frequency offset configuration. The site hopping rate in the 1D alternating configuration increases slightly with lattice spatial period, indicating somewhat less efficient damping of higher-energy excursions at longer lattice spatial periods.

There is modest variation in the temperature and site hopping rate when the cycle time of the alternating 1D configuration (Figure \ref{VarFlash1D}) and the frequency offset of the 3D configuration (Figure \ref{VarFlash3D}) are varied (these parameters are defined in the caption of Table \ref{ParamsEx}). The cooling efficiency decreases if the flash time in the 1D alternating configuration is too short; this is to be expected, as the initial transient optical pumping period after the beams switch direction, during which the atoms are not cooled efficiently, takes a greater proportion of the total time when the flash cycle time is reduced. A similar effect occurs in the 3D offset configuration; if the frequency offset between the sets of beams is too great, the change in the nature of the cooling light at the lattice site can become fast enough to adversely affect the cooling dynamics. On the other hand, if the frequency offset (in the 3D configuration) is too small, the lattice site is exposed to the molasses light with the less efficient cooling phases for a longer time each cycle; this encourages excursions to higher energies, and results in higher temperatures and decreased confinement.

The effect of the pinning lattice frequency detuning was also investigated at constant lattice depth. In the direction of higher frequency positive (blue) frequency detuning there were only minor changes in the temperatures and site lifetimes (see Table \ref{VariousRuns} for $\Delta_{l}\,$=$\,$+8000$\Gamma$). Lower positive frequency detunings were not studied due to the presence of features associated with near-resonant $F\,$=$\,1\,$$\rightarrow\,$$F'\,$=$\,$1,2$\,$($J'\,$=$\,1/2$) transitions, which are not included in the model. 

A pinning lattice with negative (red) frequency detuning was also studied; the cooling frequency detuning was altered to account for the change in the energy of an atom at the lattice minimum. Although one may expect less efficient cooling due to the atom spending more time in positions with a greater energy differential between the ground sublevels, this in fact is not the case for the lattice studied (see Table \ref{VariousRuns}). This is because the three-dimensional lattice studied has the linear polarizations of each constituent one-dimensional lattice orientated orthogonally, and consequently the energy differential between the ground sublevels cancels at the lattice potential minima. As a result, the temperatures and site lifetimes were not substantially different to the blue detuned case. However, the use of a red detuned lattice has a number of undesirable effects, such a substantial increase in the rate of spontaneous scattering of the lattice light, which increases the rate of unwanted transitions to the lower hyperfine state.

The types of lattices used in the experiments described in articles \cite{bakr_probing_2010} and \cite{sherson_single_2010} were also studied for the cooling parameters listed in Table \ref{ParamsEx} (apart for the cooling detuning, which was altered to compensate for the effects of the red-detuned lattice). The results of the simulations (see Table \ref{VariousRuns}) are similar to those quoted for the lattice of Section \ref{ExampDyn}.

\begin{table*}[t]
\centering
\caption{Temperatures and site lifetimes for different lattice frequency detunings and spatial periods. The data in this table were taken (as with Figure \ref{VarStuff}) for constant phases of the cooling beams. The quoted uncertainties are the sample standard deviations.}
\begin{tabular}{ccccc}
\hline \hline
Configuration& \multicolumn{2}{c}{1D alternating} & \multicolumn{2}{c}{3D offset} \\
Parameters & T [\textmu K] & log$_{10}$(Lifetime [s]) & T [\textmu K] & log$_{10}$(Lifetime [s])\\
\hline
$\Delta_{l}\,$=$\,$+8000$\Gamma$& 8.3$\pm$0.3 & 6.0$\pm$0.2 & 8.3$\pm$0.2 & 7.0$\pm$1.0\\ 
$\Delta_{l}\,$=$\,$-2000$\Gamma$& 8.1$\pm$0.1 & 5.6$\pm$0.2 & 7.7$\pm$0.3 & 8.3$\pm$0.7\\ 
$\Delta_{l}\,$=$\,$+8200$\Gamma$, period$\,=\,$680$\,$nm (\cite{bakr_probing_2010})& 8.7$\pm$0.2 & 6.0$\pm$0.2 & 7.1$\pm$0.2 & 7.7$\pm$1.1\\ 
1064$\,$nm counter-propagating (\cite{sherson_single_2010}) & 8.4$\pm$0.2 & 6.0$\pm$0.4 & 6.8$\pm$0.2 & 6.7$\pm$1.4\\
\hline \hline
\end{tabular}\label{VariousRuns}
\end{table*}

\subsection{Comparison of 1D alternating and 3D offset cooling methods}
\label{1D3D}

The results of the simulations predict that both the 1D alternating and the 3D offset cooling configurations produce excellent localization of a strongly-scattering atom at a pinning 
lattice site. However, the 3D offset configuration has a greater spread in the calculated temperatures, site hopping rates and photons scattering rates than the 1D alternating configuration (compare the sample standard deviations in Table \ref{ParamsOut}). 

This variability in the 3D offset configuration is primarily due to the role of the five relative phases of the cooling beams. The [0,$\,\pm\delta$] frequency detunings of the three sets of beams periodically vary the character of the cooling light at each pinning lattice site, but even so only samples a small portion of this 5-dimensional phase space, as only a single linearly independent phase parameter being varied in a cycle of frequency $\delta$. This means that changing the value of the four residual independent phases of the cooling beams does change the overall character of the light field at the lattice site, and this still affects the efficiency of the cooling at the lattice site, albeit to a lesser extent than if no frequency offset was used. 

To illustrate the residual variation in temperature and site lifetime, the worst confining set of phases for the 3D offset configuration were further investigated from the data set investigated in Section \ref{ExampDyn} (which comprises 100 runs of the HMCME simulation at randomly chosen initial phases). The temperature and site lifetime for this set of phases were found to be 9.8$\pm$0.4$\,$\textmu K and 10$^{\wedge}$(2.2$\pm$0.3)$\,$s, showing substantially less confinement than the mean. 

The photon scattering rate is also a function of these relative phases, as can be seen in the standard deviation of the scattering rate (Table \ref{ParamsOut}), which is 8\% of the mean value; this variation is not good when performing accurate fluorescent measurements. The variation in the cooling and photon scattering can be suppressed using a more complex set of frequency detunings between the cooling beams; however, there is still likely to be some residual variation between lattice sites. 

It is therefore concluded that for the situations that were simulated, while the 1D alternating and 3D offset configurations give similar temperatures and site hopping rates, the 1D alternating configuration exhibits less variability between different lattice sites and cooling laser phase arrangements, and hence is more suitable for use for fluorescent measurements.

\section{Discussion}
\label{discuss}

It is seen that for a wide range of parameters the loss rate from a site of the pinning lattice due to radiative processes can be made to be $10^{-2}\,$s$^{-1}$ or lower. The loss rate of atoms from the whole lattice due to radiative processes will occur at a lower rate: to be lost from the lattice the atom would either need to encounter another atom and suffer an inelastic light-assisted collision, or travel to the edge of a lattice without being recaptured. Even so, the transfer of atoms between wells of the pinning lattice is unwanted as it acts to blur the spatial atomic distribution measurement.

Regardless of the radiation scattering processes, the lifetime of atoms in the lattice is limited by collisions with room-temperature background gas particles, which inevitably lead to the loss of at least one atom from the lattice. The lifetime due to background collisions varies by experiment and species, but typically is around $10^{2}\,$s. If the lifetime due to radiation scattering processes is greater than around $10^{3}\,$s, the loss rate due to such processes will be negligible compared to the loss rate due to background gas collisions, and it can be said that the polarization gradient cooling is optimal.

In comparison, a typical image exposure time used in experiments is around 0.5$\,$s to 1$\,$s \cite{nelson_imaging_2007,bakr_quantum_2009,bakr_probing_2010,sherson_single_2010}. Nevertheless, it is in general desirable to have the site lifetime as long as possible, which in practice means that it is limited by background gas collisions. This is because the site lifetime has a direct bearing on the fidelity of the measurement: for a 1$\,$s measurement, 1\% of atoms are lost during the measurement if they have a 100$\,$s lifetime, but around 10\% are lost for a 10$\,$s lifetime; this is significant if hundreds of atoms are present in the image. Furthermore, and more perhaps importantly for future experiments, three-dimensional tomographic measurements have longer exposure times than two-dimensional measurements, as more images need to be taken; consequently it is more critical that the site lifetime is as long as possible.

In fact the results show that the radiative site lifetime is several orders of magnitude above $10^{3}\,$s for a wide range of parameters. This may be important for experiments as it adds to the robustness of the measurement technique. Consider that an additional source of heating is present, for example due to the reabsorption of scattered radiation; the increased heating will tend to diminish the site lifetime exponentially. If the parameters of the cooling light are chosen so there is a `safety margin' between the radiative site lifetime and the background gas collisional lifetime, the measurement technique can be made robust to the extra heating. 

Another very important consideration is that in an experiment it may be useful to use additional laser beams, on top of the ones discussed in this article, in order to aid measurements on the sample. An example of this concept is discussed in a separate article \cite{shotter_design_2010}, where it is proposed to use an additional laser beam to be able to distinguish between atoms undergoing fluorescent scatter at different depths in a three-dimensional sample; this would be of clear benefit for tomographic measurements. The additional probing beam adds substantial extra heating to the atoms, and it is beneficial to use near-optimal cooling parameters in order to diminish atom losses during the measurement.

\section{Conclusion}

The polarization gradient cooling of atoms trapped in a deep off-resonant optical lattice was simulated using a hybrid Monte Carlo-Master Equation technique. The calculations indicate that there is a wide parameter range in which the lifetime at a lattice site of a strongly-scattering single atom undergoing polarization gradient cooling is primarily limited by collisions with background gas. This is consistent with recent experiments \cite{nelson_imaging_2007,bakr_quantum_2009,bakr_probing_2010,sherson_single_2010}. 

The calculations suggest that for near-optimal parameters the radiative loss rate from a site is in fact orders of magnitudes less than the loss rate due to background gas collisions. This could be important for experiments, as it suggests that the technique can be made robust in regards to additional heating mechanisms not included in the present analysis, for example heating from rescattered light or from additional probing beams.

It has been shown that by the use of an in-lattice polarization gradient cooling fluorescent measurement technique it is possible to extract large photon numbers of up to $10^{8}$ from each atom during fluorescent measurements of a dilute atomic sample. This strongly supports the use of this technique when probing the spatial distribution of low atom-number samples of ultracold atoms.
The author would like thank Christopher Foot (Oxford, UK), members of the Quantum Processes and Metrology Group (NIST, Gaithersburg, USA), and the NIST internal referees for their comments, and to acknowledge funding from the EPSRC (UK), Christ Church College, Oxford (UK), The Lindemann Trust (UK) and the Atomic Physics Division of NIST, Gaithersburg (USA). 

\begin{appendix}
\section{The angular distribution of spontaneously emitted radiation} 
\label{dir_sp}

The intensity of the spontaneously scattered radiation is (\cite{loudon_quantum_2000} p.173)
\begin{equation}\label{closc2}
I({\bm r}) \propto \left \langle {\bm E^{-}({\bm r})} {\bm E^{+}}({\bm r}) \right \rangle \; .
\end{equation}
The electric field of the scattered light in the far field is determined from the source-field expression (\cite{loudon_quantum_2000} p.328)
\begin{equation}\label{closc3}
{\bm E^{+}}_{\alpha}({\bm r}) \propto {\bm e}_{\alpha} \left ({\bm e}_{\alpha} \cdot {\bm D^{\bm -}}\right ) \; 
\end{equation}
in which the unit vector ${\bm e}_{\alpha}$ is the polarization vector of the radiation mode labeled by $\alpha$.

Dividing the intensity into two perpendicular polarizations, conveniently $ \hat{\bm \theta}$ and $ \hat{\bm \phi}$ in spherical polar coordinates, the following relationships are obtained:
\begin{eqnarray}\label{closc4}
I_{\theta} ({\bm r})&\propto& \left \langle ( \hat{\bm \theta}({\bm \kappa}) \cdot {\bm D^{\bm +}} ) \; ( \hat{\bm \theta}({\bm \kappa}) \cdot {\bm D^{\bm -}}) \right \rangle \\
I_{\phi}({\bm r}) &\propto& \left \langle (\hat{\bm \phi}({\bm \kappa}) \cdot {\bm D^{\bm +}}) \; (\hat{\bm \phi}({\bm \kappa}) \cdot {\bm D^{\bm -}})\right \rangle
\end{eqnarray}
in which mode $\alpha$ has been identified by the unit wave vector ${\bm \kappa}$ and the polarization. The expectation value runs over the internal degrees of freedom of the atom. By breaking down the dipole operator into the component basis, the intensity is expressed in the form
\begin{equation}\label{closc6}
I_{\sigma}({\bm r}) = \frac{\hbar \omega \Gamma}{r^{2}} \sum_{ \varepsilon \varepsilon'} f^{\sigma}_{\varepsilon \varepsilon'} \left \langle {\bm D^{\bm +}_{\bm \varepsilon}}{\bm D^{\bm -}_{\bm \varepsilon'}} \right \rangle \; .
\end{equation}
The coefficients $f^{\sigma}_{\varepsilon \varepsilon'}$ are listed in Table \ref{f_params} with the normalization chosen so that the total scattered power is $\hbar \omega \Gamma \langle {\bm P} \rangle\;$. 

The tensor $\eta_{\varepsilon \varepsilon' i j}$ of Equation \ref{wig_app} is related to the functions $f^{\sigma}_{\varepsilon \varepsilon'}$ by
\begin{equation}\label{closc7}
\eta_{\varepsilon \varepsilon' i j}=\sum_{\sigma} \int d^{2}{\bm \kappa} ({\bm \kappa} \cdot {\bm \hat{x}_{i}})( {\bm \kappa} \cdot {\bm \hat{x}_{j}}) f^{\sigma}_{\varepsilon \varepsilon'} 
\end{equation}
with ${\bm \hat{x}_{i} }$ being the unit vector in the $i$ direction. The tensor $\eta_{\varepsilon \varepsilon' i j}$ is given in Table \ref{eta}.

\begin{table*}
\centering
\caption{The parameters $f^{\sigma}_{\varepsilon \varepsilon'}$ of Equation \ref{sp_dyn_rec} in terms of the polar angle $\theta$ and azimuthal angle $\phi$. }
\newcolumntype{A}{ >{\centering\arraybackslash$\displaystyle}m{1.4in} <{$ } }
\newcolumntype{B}{ >{\centering\arraybackslash$\displaystyle}m{0.7in} <{$ } }
\begin{center}
\begin{tabular}{B|AAA}
\rule[-10pt]{0pt}{25pt} f^{\theta}_{\varepsilon \varepsilon'}&\rule[-9pt]{0pt}{25pt} \varepsilon'=-1&\rule[-9pt]{0pt}{25pt} \varepsilon'=0&\rule[-9pt]{0pt}{25pt} \varepsilon'=1
\\ \hline \rule[-17pt]{0pt}{40pt} 
\varepsilon=-1&\rule[-17pt]{0pt}{40pt}\frac{3}{16\pi} \cos^{2} \theta &\rule[-17pt]{0pt}{40pt}-{\frac {3}{16\sqrt {2}\pi}} \sin 2\theta {e^{-i\phi}}&\rule[-17pt]{0pt}{40pt} -\frac{3}{16\pi} \cos^{2} \theta {e^{-2i\phi}} \\ \rule[-17pt]{0pt}{40pt}
\varepsilon=0&\rule[-17pt]{0pt}{40pt}-{\frac {3}{16\sqrt {2}\pi}} \sin 2\theta {e^{i\phi}}&\rule[-17pt]{0pt}{40pt}\frac{3}{8\pi} \sin^{2} \theta &\rule[-17pt]{0pt}{40pt} {\frac {3}{16\sqrt {2}\pi}} \sin 2\theta {e^{-i\phi}} \\ \rule[-17pt]{0pt}{40pt} 
\varepsilon=1&\rule[-17pt]{0pt}{40pt}-\frac{3}{16\pi} \cos^{2} \theta {e^{2i\phi}}&\rule[-17pt]{0pt}{40pt}{\frac {3}{16\sqrt {2}\pi}} \sin 2\theta {e^{i\phi}}&\rule[-17pt]{0pt}{40pt} \frac{3}{16\pi} \cos^{2} \theta \\ 
\end{tabular}
\end{center}
\centering
\begin{tabular}{B|AAA}
\rule[-10pt]{0pt}{25pt} f^{\phi}_{\varepsilon \varepsilon'}&\rule[-9pt]{0pt}{25pt} \varepsilon'=-1&\rule[-9pt]{0pt}{25pt} \varepsilon'=0&\rule[-9pt]{0pt}{25pt} \varepsilon'=1
\\ \hline \rule[-17pt]{0pt}{40pt} 
\varepsilon=-1&\rule[-17pt]{0pt}{40pt}\frac{3}{16\pi} &\rule[-17pt]{0pt}{40pt} 0 &\rule[-17pt]{0pt}{40pt} \frac{3}{16\pi} {e^{-2i\phi}} \\ \rule[-17pt]{0pt}{40pt}
\varepsilon=0&\rule[-17pt]{0pt}{40pt}0&\rule[-17pt]{0pt}{40pt}0&\rule[-17pt]{0pt}{40pt}0\\ \rule[-17pt]{0pt}{40pt} 
\varepsilon=1&\rule[-17pt]{0pt}{40pt}\frac{3}{16\pi} {e^{2i\phi}}&\rule[-17pt]{0pt}{40pt}0&\rule[-17pt]{0pt}{40pt} \frac{3}{16\pi} \\ 
\end{tabular}
\label{f_params}
\end{table*}
\begin{table*}
\centering
\caption{The tensor $\eta_{\varepsilon \varepsilon' i j}$ of Equation \ref{wig_app}. }
\label{eta}
\newcolumntype{A}{ >{\raggedleft\arraybackslash$\displaystyle}m{1.6in} <{$ } }
\newcolumntype{B}{ >{\raggedleft\arraybackslash$\displaystyle}m{0.55in} <{$ } }
\begin{tabular}{B|AAA}
\rule[-10pt]{0pt}{25pt} \eta_{\varepsilon \varepsilon' i j}& \rule[-9pt]{0pt}{25pt} \varepsilon'=-1$\hspace{20pt}$& \rule[-9pt]{0pt}{25pt} \varepsilon'=0$\hspace{20pt}$&\rule[-9pt]{0pt}{25pt} \varepsilon'=1$\hspace{20pt}$
\\ \hline \rule[-17pt]{0pt}{40pt} 
\varepsilon=-1\hspace{10pt}
&\rule[-25pt]{0pt}{60pt}\frac{1}{10}\left[\begin{array}{ccc}3&0&0\\ 0&3&0\\ 0&0&4\end{array}\right] 
&\rule[-25pt]{0pt}{60pt}\frac{1}{10\sqrt{2}}\left[\begin{array}{ccc}0&0&-1\\ 0&0&i\\ -1&i&0 \end{array}\right] 
&\rule[-25pt]{0pt}{60pt}\frac{1}{10}\left[\begin{array}{ccc}1&-i&0\\ -i&-1&0\\ 0&0&0 \end{array}\right] 
\\ \varepsilon=0\hspace{10pt}
&\rule[-25pt]{0pt}{60pt}\frac{1}{10\sqrt{2}}\left[\begin{array}{ccc}0&0&-1\\ 0&0&-i\\ -1&-i&0 \end{array}\right] 
&\rule[-25pt]{0pt}{60pt}\frac{1}{5}\left[\begin{array}{ccc}2&0&0\\ 0&2&0\\ 0&0&1 \end{array}\right] 
&\rule[-25pt]{0pt}{60pt}\frac{1}{10\sqrt{2}}\left[\begin{array}{ccc}0&0&1\\ 0&0&-i\\ 1&-i&0 \end{array}\right] 
\\ \varepsilon=1\hspace{10pt}
&\rule[-25pt]{0pt}{60pt}\frac{1}{10}\left[\begin{array}{ccc}1&i&0\\ i&-1&0\\ 0&0&0 \end{array}\right] 
&\rule[-25pt]{0pt}{60pt}\frac{1}{10\sqrt{2}}\left[\begin{array}{ccc}0&0&1\\ 0&0&i\\ 1&i&0 \end{array}\right] 
&\rule[-25pt]{0pt}{60pt}\frac{1}{10}\left[\begin{array}{ccc}3&0&0\\ 0&3&0\\ 0&0&4 \end{array}\right] 
\\
\end{tabular}
\end{table*}
\pagebreak

\section{The Wigner transformation}
\label{spont_append}

The transformation of the kinetic energy operator term of the unitary dynamics (Equation \ref{main_ham}) is covered in standard treatments of the derivation of the Wigner-Moyal equation \cite{schleich_quantum_2001,liboff_kinetic_2003}. The transformation of the potential energy operator term of Equation \ref{main_ham} proceeds straightforwardly using the position space representation of the Wigner transformation (Equation \ref{wiga}). For the non-unitary dynamics of Equation \ref{sp_dyn_rec}, the transformation of the terms containing the constant matrix ${\bm P}$ is trivial. The remaining term of the equation is
\begin{equation} \label{sp_dyn_rec2}
\Gamma \sum_{\varepsilon \varepsilon' \sigma} \int d^{2}{\bm \kappa} \: e^{i k_{R}{\bm \kappa}.{\bm r'}} \: {\bm D^{\bm -}_{\bm \varepsilon}}{\bm \rho}{\bm D^{\bm +}_{\bm \varepsilon'}} \: e^{-i k_{R}{\bm \kappa}.{\bm r}} \: f^{\sigma}_{\varepsilon \varepsilon'}({\bm \kappa}) \; .
\end{equation}
For this term it is easier to use the Wigner transformation as expressed in the momentum representation, which is found to be
\begin{equation} \label{wig2}
\mathcal{W}\hat{\bm A}({\bm p},t)=\frac{1}{h^3} \int d^{3}{\bm q}\left\langle {\bm
p}+\frac{\bm q}{2} \right| \hat{\bm A} \left| {\bm p}-\frac{\bm q}{2}
\right\rangle e^{ i {\bm r}\cdot {\bm q}/\hbar}\; .
\end{equation}
The application of this transform onto the Expression \ref{sp_dyn_rec2} involves expressions such as 
\begin{equation} \label{wig3}
e^{-i k_{R}{\bm \kappa}.{\bm r}}|{\bm p}\rangle = |{\bm p}-\hbar k_{R} {\bm \kappa}\rangle ,
\end{equation}
which are evaluated by identification of the momentum translation operator. The matrices ${\bm D^{\bm \pm}_{\bm \varepsilon}}$ are constant in spatial coordinates, so the Wigner transformed term is
\begin{equation}
\Gamma \sum_{\varepsilon \varepsilon' \sigma} \int d^{2}{\bm \kappa} {\bm D^{\bm -}_{\bm \varepsilon}}{\bm W}({\bm r},{\bm p}-\hbar k_{R} {\bm \kappa},t){\bm D^{\bm +}_{\bm \varepsilon'} } f^{\sigma}_{\varepsilon \varepsilon'}({\bm \kappa}) \; .
\end{equation}

\section{The semiclassical approximation and conversion to Langevin form}
\label{ham_append}

The semiclassical approximation is taken by the termination of the expansion (Equation \ref{wig_exp}) of the right hand side of Equation \ref{wig_comp} in terms of small changes in momenta. Terms containing polynomials in ${\bm s}$ are expressed in terms of the derivative of the Wigner function with respect to momentum, for example:
\begin{equation} \label{wigc}
\nabla_{p} \int d^{3}{\bm s}\left\langle {\bm
r}+\frac{\bm s}{2} \right| {\bm \rho} \left| {\bm r}-\frac{\bm s}{2}
\right\rangle e^{- i {\bm p}\cdot {\bm s}/\hbar}=-\frac{i}{\hbar}\int d^{3}{\bm s} \left\langle {\bm
r}+\frac{\bm s}{2} \right| {\bm \rho} \left| {\bm r}-\frac{\bm s}{2} 
\right\rangle e^{- i {\bm p}\cdot {\bm s}/\hbar} {\bm s} \; .
\end{equation}

The recoil term of Equation \ref{wig_comp} is directly expanded in terms of small momenta:
\begin{equation} \label{wig_exp2}
{\bm W}({\bm r},{\bm p}+{\bm q'},t)= {\bm W}({\bm r},{\bm p},t) + {\bm q'}\cdot \nabla_{p}{\bm W}({\bm r},{\bm p},t)+ \frac{1}{2} \left ( {\bm q'}\cdot \nabla_{p}\right )^{2}{\bm W}({\bm r},{\bm p},t)+\ldots \; .
\end{equation}
Integrals are performed over the unit sphere, and a sum taken over polarizations. Terms containing odd powers of kappa vanish, leaving a residual second order term 
\begin{equation} \label{wig_exp3}
\frac{\Gamma \hbar^{2}k_{R}^{2}}{2}\sum_{\varepsilon,\varepsilon',i,j} \eta_{\varepsilon \varepsilon' i j}{\bm D^{\bm -}_{\bm \varepsilon}}\frac{\partial^{2} {\bm W}}{\partial p_{i}\partial p_{j}}{\bm D^{\bm +}_{\bm \varepsilon'} } \; .
\end{equation}
The tensor $\eta_{\varepsilon \varepsilon' i j}$ defined in Equation \ref{closc7}.

The conversion of Equation \ref{wig_app} to Langevin form is achieved by substitution of the trial solution of Equation \ref{delta_app}. To find the equation of motion for the internal coordinates (Equation \ref{internal_app}) the substitution is performed directly and the integral taken over the external co-ordinates. To find the equations of motion for the external co-ordinates (Equations \ref{motion} to \ref{motion_end}) the trial solution is substituted and the appropriate expectation value taken (by integration over both internal and external coordinates). A key relation concerns the derivatives of the delta distribution
\begin{equation} \label{delta_deriv}
\int dx f(x) \frac{\partial^{n} \delta}{\partial x^{n}}=-\int dx \frac{df(x)}{dx} \frac{\partial^{n-1} \delta}{\partial x^{n-1}}\; .
\end{equation}
Using this relation the equations of motion are found for $\tilde{\bm r}=\langle{\bm r}\rangle$ etc. The non-zero expectation values for the second moments of momentum are interpreted as a Langevin force.

\section{Summary of Hybrid Monte Carlo-Master Equation method}
\label{HMCME}

A measure is chosen which is representative of the dynamics of interest (see Section \ref{measEv}), and a series of points are chosen on this measure. The value of the measure varies in time as the system evolves. The system is classified at any time according to the last point encountered by the measure. When the system's classification changes (when the measure encounters a new point), the state vector of the system is recorded; this is called a `start vector' for the new point.
\begin{enumerate}
\item \begin{enumerate} \item One or more initial standard Monte Carlo simulations are carried out. After the system settles into a steady state, the state vector is recorded every time the system encounters a new point; these form the initial population of start vectors for the rest of the HMCME simulation. This continues until enough start vectors have been found to characterize a few of the points on the measure (those where the system is most commonly found). 
\item An equal weight is assigned to each of the initial start vectors.
\end{enumerate}
\item \begin{enumerate} \item A single start vector (the mother) is chosen randomly from the set of all start vectors at partially filled points with a probability which is proportional to the start vector's weight $w_{mth}$.
\item The mother start vector is propagated multiple times using the Monte Carlo method until the next point is reached. The resulting state vectors (the daughters) are categorized by end point; a single randomly chosen daughter vector is retained for each end point. The branching probabilities $p_{ij}$ for the mother start vector are calculated.
\item Weights $\left(w_{dt}\right)_{j}\,$$=\,$$p_{ij}w_{mth}$ are assigned to the two daughter vectors. If one of the daughter vectors is at a full point, the daughter's weight is added to the combined weight at that point.
\item For each full point (with weight $W_{j}$) a new weight is assigned according to
\begin{equation} \label{D1}
\left(W_{j}\right)_{new}=W_{j}-\frac{w_{mth}W_{j}}{W_{N}}\;
\end{equation}
with ${W_{N}}$ the total weight at all partially filled points. The weight from this point is distributed to the two neighboring points. If a neighboring point (with weight $W_{k}$) is also full, it is assigned the new weight
\begin{equation}
\left(W_{k}\right)_{new}=W_{k}+\frac{p_{jk}w_{mth}W_{j}}{W_{N}}.
\end{equation}
where $p_{jk}$ are the calculated mean branching ratios from $j$ to $k$. If the neighboring point $k$ is not full the start vectors at that point which originate from point $j$ are assigned the weights
\begin{equation} \label{D3}
\left(w_{dt\,j\rightarrow k}\right)_{new}=w_{dt\,j\rightarrow k}+\frac{p_{jk}w_{mth}W_{j}}{W_{N}}\frac{w_{dt\,j\rightarrow k}}{\sum w_{dt\,j\rightarrow k}}
\end{equation}
where the summation $\sum w_{dt\,j\rightarrow k}$ runs over all the daughter vectors of the point $j$ at point $k$. The reallocation of weights described in this step is carried out so weight is reallocated from each full point.
\item A weight $w_{mth}\,$$=\,$$0$ is assigned to the mother vector.
\item If the number of start vectors propagated reaches the maximum allowed $M$ at a point $i$, the point is called `full'; the weights $w_{i,n}$ of all start vectors at this point are added to give $W_{i}=\sum_{n}w_{i,n}$, and the mean branching ratios $p_{ij}$ are found for all relevant channels.
\item Part 2 is repeated until data has been taken for all points of interest.
\end{enumerate}
\item The results of the Part 2 are processed to form a master equation as discussed in Section \ref{MarkNonMark}.
\end{enumerate}

The reasoning behind the reallocation of weights from the full points (Step 2(d)) is to prevent a false buildup of weights at the full points; in the absence of this reallocation, weight is transferred from partially filled to full points, but not the other way around. To compensate for not propagating vectors from these full points, the weights are reallocated (Equations \ref{D1} to \ref{D3}) in the same proportion that would occur if start vectors were propagated from the full points. These equations are derived by noting that, when each start vector is propagated, a fraction $w_{mth}/W_{N}$ of the weight of $W_{N}$ (the total weight at partially filled points, i.e. the total weight from which the start vector is chosen) is being reallocated; therefore, to compensate for not propagating start vectors from the full points, an equivalent fraction of the weight at full points $W_{j}\times (w_{mth}/W_{N})$, for all full points $j$, should be reallocated according to the known branching ratios.
\end{appendix}

\end{document}